\documentclass[twocolumn,10pt,pra,aps,showpacs]{revtex4-1}

\usepackage{graphicx}
\usepackage{amsbsy}
\usepackage{amsmath}
\usepackage{subfigure}
\usepackage[utf8x]{inputenc}
\usepackage{natbib}

\newcommand{\pa}[2]{\frac{\partial #1}{\partial #2}}

\begin{document}
\title{Photon-pair generation in nonlinear
metal-dielectric 1D photonic structures}
\author{D. Jav\r{u}rek}
\email{javurek@slo.upol.cz}
\address{RCPTM, Joint Laboratory of Optics of Palack\'y University
and Institute of Physics of AS CR, 17. listopadu 12, 771 46 Olomouc,
Czech Republic}
\author{J. Pe\v{r}ina Jr.}
\address{Institute of Physics, Joint Laboratory of Optics of Palack\'y
University and Institute of Physics of AS CR, 17. listopadu 50a,
771 46 Olomouc, Czech Republic}
\author{J. Svozil\'{i}k}
\address{RCPTM, Joint Laboratory of Optics of Palack\'y
University and Institute of Physics of AS CR, 17. listopadu 12,
771 46 Olomouc, Czech Republic}

\begin{abstract}
Nonlinear metal-dielectric layered structures are shown to be able
to efficiently generate entangled photon pairs using spontaneous
parametric down-conversion. Increase of electric-field amplitudes
in these structures enhanced by the presence of metal layers is
sufficient to compensate for losses inside thin metal layers. As
an example, photon pairs emitted from a structure composed of
alternating nonlinear dielectric GaN layers and metal Ag layers
are analyzed in spectral, temporal as well as spatial domains.
Also correlations and entanglement between two photons in a pair
are determined. Very narrow photon-pair spectra together with
strong directionality of photon-pair emission are observed making
the photons suitable for photon-atom interactions. Highly enhanced
electric-field amplitudes provide high photon-pair generation
efficiencies.
\end{abstract}

 \maketitle


\section{Introduction}

Spontaneous parametric down-conversion (SPDC) is a quantum
nonlinear process that was predicted in 1961 \cite{Louisell1961}
and experimentally observed for the first time in 1968
\cite{Harris1967,Magde1967}. SPDC occurs in nonlinear media with
nonzero second-order nonlinear susceptibility tensor $\chi^{(2)}$.
During this process, the conservation law of energy originating in
homogeneity of time is fulfilled. Also the conservation law of
momentum is usually obeyed, at least for the transverse components
of wave vectors of the interacting fields. This law originating in
homogeneity of space is approximately valid also for longer
homogeneous crystals along the propagation direction. The
generation of photon pairs has to fulfill both laws and so photon
pairs typically occur in states entangled in frequencies, momenta,
orbital angular momenta or polarizations
\cite{Keller1997,Svozilik2011,Svozilik2012,Grice2008,Law2004}.

Phase-matching conditions can only be fulfilled under specific
conditions that determine the properties of photon pairs. For this
reason, new and efficient sources of photon pairs have been
developed using, e.g., periodically poled crystals. Periodical
poling which introduces periodical modulation of $ \chi^{(2)} $
nonlinearity offers enhanced control of phase matching of the
nonlinear process as well as modification of spectral properties
of the emitted photon pairs
\cite{Harris2007,Brida2009,Svozilik2009}.

Modern optical structures that confine the fields in one (layered
structures) or two (waveguides, optical fibers) dimensions
represent qualitative improvement from the point of view of
efficiency of photon-pair generation. The confinement of
interacting fields enhances their electric-field amplitudes on one
side, it qualitatively changes the conditions for an efficient
nonlinear interaction on the other side. The requirement for phase
matching of wave vectors is then replaced by the need of large
spatial overlap of the electric-field amplitudes of all three
interacting fields. As spatial profiles of the electric-field
amplitudes depend strongly on parameters and geometry of the
structures, much broader possibilities for tailoring properties of
the emitted photon pairs exist.

A great deal of attention has been devoted to waveguiding
structures including planar or rectangular waveguides and photonic
fibers. Two-dimensional confinement, together with sufficiently
long structures, provides high absolute conversion efficiencies of
SPDC, even three or four orders in magnitude larger compared to
typical nonlinear bulk crystals. Efficient SPDC in
periodically-poled waveguides has been investigated in
\cite{Eckstein2011,Jachura2014,Machulka2013}. On the other hand,
SPDC in photonic fibers \cite{Ren2004, Sang2006, Zhu2012} provides
photon pairs in transverse (guided) modes with radial symmetry,
that are pivotal for optical-fiber communications. From the
perspective of applications in communications, ring and vortex
nonlinear silica fibers are promising
\cite{Javurek2014a,Javurek2014b}.

As already mentioned, nonlinear layered structures confine the
fields along their propagation direction. Back-scattering of the
fields creates a one dimensional photonic-band structure (PBG)
with transmission peaks and forbidden bands
\cite{Scalora1997,Centini2005,PerinaJr2006,PerinaJr2011}. The
electric-field amplitudes are enhanced by this back-scattering,
which under suitable conditions gives an efficient nonlinear
interaction. However, as the confinement of optical fields occurs
only in one dimension, the enhancement of optical fields is
considerably weaker compared to waveguiding structures, at least
for dielectric structures. On the other hand, there exist the
usual transverse phase-matching conditions and the impinging
fields can be easily coupled into the modes of the structure
\cite{PerinaJr2011}. Also properties of a two-photon state can be
efficiently and easily controlled by spatial and temporal spectra
of the pump beam. Taking into account the precision of
well-established fabrication techniques, one-dimensional PBGs
represent promising sources of photon pairs.

Nonlinear dielectric layered structures have been already
investigated from the point of view of SPDC. Both semiclassical
(stochastic) \cite{Centini2005} and quantum models
\cite{PerinaJr2006,PerinaJr2011} of SPDC in dielectric layered
structures have been elaborated. These structures have been shown
to be able to provide entangled photon pairs anti-symmetric with
respect to the exchange of signal and idler frequencies
\cite{PerinaJr2007b}. Also random nonlinear dielectric layered
structures have been analyzed as sources of spectrally
ultra-narrow photon pairs \cite{PerinaJr2009b,PerinaJr2009c}.
Surface SPDC has been shown to give important contribution to
photon-pair generation rates
\cite{PerinaJr2009,PerinaJr2009a,Perinova2013}.

On the contrary, metal-dielectric layered structures have been
investigated from the point of view of transmission properties
\cite{Scalora1999,Gadson2009}. It has been shown that, considering
the overall transmission, the total amount of metal inside the
structure can be considerably larger provided that it is split
into thin layers sandwiched by dielectric layers. This occurs due
to strong back-scattering on metal-dielectric boundaries with high
contrast of refraction indices. This contrast is not only
sufficient for the compensation of losses in metal layers, it also
enhances the electric-field amplitudes considerably stronger
compared to only dielectric structures \cite{PerinaJr2014}. It
also allows us to consider efficient nonlinear processes in more
complex metal-dielectric structures. Narrow spectral interaction
regions and strong directionality of photon emissions are
distinguished properties of such structures. For this reason, the
emitted photon pairs are suitable for photon-atom interactions
that require both properties to maximize the strength of
interaction \cite{Kolchin2006}. We note that such photon-atom
interaction is in the center of attention in recent years in
quantum-information processing as entanglement is easily generated
in optical fields but excitations are easily stored in atomic
systems \cite{Raimond2011,Volz2006,Choi2008}. Recently, the
process of second harmonic generation in metal-dielectric layered
structures has been investigated both theoretically and
experimentally \cite{Lepeshkin2004, Scalora2010}. Also the first
brief investigation of SPDC in such structures has confirmed high
enhancement of photon-pair generation rates due to strong
back-scattering occurring at metal-dielectric boundaries with high
contrast of refraction indices \cite{Javurek2012}. In the paper,
we extend this investigation to provide a comprehensive study of
properties of photon pairs emitted in metal-dielectric layered
structures.

Optical nonlinear response of metals can arise due to several
physical processes including the Fermi smearing
\cite{Lepeshkin2004}, strong redistribution of charges
\cite{Scalora2010,Ginzburg2010} and affecting the path of
electrons by a strong magnetic field. Other mechanisms leading to
nonlinearity are discussed in \cite{Scalora2010,Larciprete2008}.
In this paper, we derive nonlinearity of the considered Ag layers
from the action of the Lorentz force on electrons
\cite{Larciprete2008}. As for the dielectric layers, we consider
GaN that is transparent for the pump field at wavelength $
\lambda_p = 400 $~nm and thus allows the generation of photon
pairs with wavelengths around $ \lambda = 800 $~nm efficiently
detected at single-photon level by Si-based detectors. Moreover,
GaN has sufficiently high $ \chi^{(2)} $ nonlinearity and the
fabrication of thin layered GaN structures is well mastered.

The paper is organized as follows. The model of SPDC in
metal-dielectric layered structures is presented in Sec.~II.
Physical quantities characterizing the emitted photon pairs are
described in Sec.~III. In Sec.~IV, a metal-dielectric resonator
composed of two Ag layers and one GaN layer is analyzed. An
efficient structure composed of eleven GaN and Ag layers is suggested
and analyzed as a typical example in Sec.~V. Temporal properties
of the emitted photon pairs are investigated in Sec.~VI. Noise
originating in losses in metal layers is addressed in Sec.~VII.
Conclusions are drawn in Sec.~VIII. Appendix~A brings the
derivation of $\chi^{(2)}$ tensor for metals. Extension of the
theory quantifying the noise is given in Appendix~B.

\section{Model of spontaneous parametric down-conversion}

Vectorial model of SPDC in nonlinear layered structures has been
formulated in \cite{PerinaJr2011} using the interaction
Hamiltonian $ \hat{H}_{\rm int} $. Alternatively, the interaction
momentum operator $ \hat{G}_{\rm int} $ can be used to describe
SPDC caused by a strong pump beam propagating along the $ z $ axis
\cite{Mandel1995,PerinaJr2000,PerinaJr2014}:
\begin{eqnarray}  
 \hat{G}_{\rm int}(z) &=& 2 \varepsilon_0 \int_{-\infty}^{\infty} dt
  \int_{\cal S} dxdy \;  \chi^{(2)}({\bf r}) \nonumber \\
 & & \mbox{} :
 \left[ {\bf E}_{p}^{(+)}({\bf r},t) \hat{\bf E}_{s}^{(-)}({\bf r},t)
 \hat{\bf E}_{i}^{(-)}({\bf r},t) + {\rm h.c.} \right];
\label{1}
\end{eqnarray}
$ {\bf r} = (x,y,z) $. The pump-field is characterized by its
positive-frequency electric-field vector amplitude $ {\bf
E}_{p}^{(+)}({\bf r},t) $. The signal and idler fields are
described by their negative-frequency electric-field operator
vector amplitudes $ \hat{\bf E}_{s}^{(-)}({\bf r},t) $ and $
\hat{\bf E}_{s}^{(-)}({\bf r},t) $, respectively. Shortening of
the tensor of nonlinear susceptibility $ \chi^{(2)} $ with respect
to its three indices is denoted by $ : $. Symbol $ \varepsilon_0 $
stands for the vacuum permittivity; $ {\rm h.c.} $ replaces the
Hermitian conjugated term. We note that whereas the nonlinear
interaction Hamiltonian $ \hat{H}_{\rm int} $ gives the
interaction energy, the momentum operator $ \hat{G}_{\rm int}(z) $
provides the overall flux of this energy through the transverse
plane $ {\cal S} $ positioned at distance $ z $.

\begin{figure} 
\centering
 \includegraphics[scale=1]{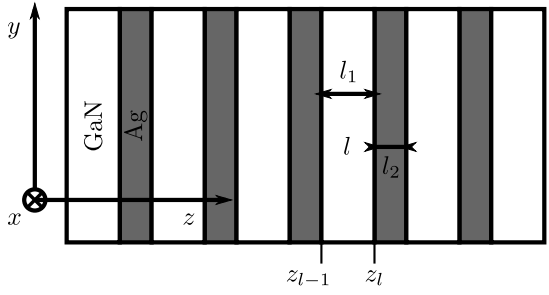}
 \caption{Scheme of a metal-dielectric layered structure composed
 of six GaN layers and five Ag layers.}
 \label{fig1}
\end{figure}

The strong un-depleted pump field is characterized by its incident
temporal spectrum $ {\cal E}_p(\omega_p) $ and spatial spectrum $
{\cal E}_p^{\rm tr}(k_{p,x},k_{p,y}) $ defined in the transverse
plane $ {\cal S} $. The pump positive-frequency amplitude $ {\bf
E}_{p}^{(+)}({\bf r},t) $ occurring in Eq.~(\ref{1}) can be
decomposed in a layered structure with boundaries localized at
positions $ z_j $, $ j=0,\ldots,N $, (for the scheme of the
structure, see Fig.~\ref{fig1}) as follows:
\begin{eqnarray}   
 {\bf E}_{p}^{(+)}({\bf r},t) &=&
  \frac{1}{\sqrt{2\pi}^3 c^2} \int_{-\pi/2}^{\pi/2}
  |\sin(\vartheta_p)| \, d\vartheta_p \int_{-\pi/2}^{\pi/2} d\psi_p
   \nonumber \\
 & & \hspace{-5mm} \int_{0}^{\infty} \omega_p^2 d\omega_p \; {\cal E}_p(\omega_p)
  {\cal E}_p^{\rm tr} \left[ k_{p,x}
  ({\bf \Omega_p}), k_{p,y}({\bf \Omega_p}) \right]  \nonumber \\
 & & \hspace{-5mm} \mbox{} \times \exp \left[ i k_{p,x}({\bf \Omega_p})
  x + i k_{p,y}({\bf \Omega_p}) y\right] \sum_{\gamma={\rm TE,TM}} \nonumber \\
 & & \hspace{-5mm} \mbox{} \sum_{g=F,B}
  \sum_{l=0}^{N+1} {\rm rect}^{(l)}(z) A_{p_g,\gamma}^{(l)}({\bf
  \Omega_p}) {\bf e}_{p,\gamma}^{(l)}({\bf \Omega_p}) \nonumber \\
 & & \hspace{-5mm} \mbox{} \times
  \exp\left[ ik_{p_g,z}^{(l)}({\bf \Omega}_p) (z-z_{l-1}) \right]
   \exp(-i\omega_p t)
\label{2}
\end{eqnarray}
using the notation $ {\bf \Omega}_p \equiv
(\omega_p,\vartheta_p,\psi_p) $ for 'spherical coordinates'
composed of the frequency $ \omega_p $, radial propagation angle
$\vartheta_p $ and azimuthal propagation angle $ \psi_p $. The
scalar electric-field amplitudes $ A_{p_F,\gamma}^{(l)} $ and $
A_{p_B,\gamma}^{(l)} $ in Eq.~(\ref{2}) characterize the forward-
and backward-propagating pump fields, respectively, with $ \gamma
$ polarization in an $ l $-th layer with index of refraction $
n_p^{(l)} $. Polarization vectors $ {\bf e}^{(l)}_{p_F,\gamma} $
and $ {\bf e}^{(l)}_{p_B,\gamma} $ determine polarization
directions of $ \gamma $-polarized waves in an $ l $-th layer
propagating forward (index $ F $) and backward ($ B $),
respectively. Function $ {\rm rect}^{(l)}(z) $ for $ l=1,\ldots,N
$ equals 1 for $ z_{l-1} \le z < z_l $ and is zero otherwise;
function $ {\rm rect}^{(0)}(z) $ [$ {\rm rect}^{(N+1)}(z) $] is
nonzero only for $ z<z_0 $ [$ z_N \le z$] and equals 1. Speed of
light in vacuum is denoted as $ c $. Decomposition of the pump
electric-field amplitude $ {\bf E}_{p}^{(+)} $ into its TE- and
TM-polarized waves \cite{Yeh1988} in Eq.~(\ref{2}) is done with
respect to the plane of incidence of a plane wave with given wave
vector $ {\bf k}_p $.

Cartesian components of the pump-field wave vector $ {\bf k}_p $
can be written in the form:
\begin{eqnarray}  
 k_{p,x}({\bf \Omega}_p) &=&
  - \frac{\omega_p\sin(\psi_p)\sin(\vartheta_p) }{c} , \nonumber\\
 k_{p,y}({\bf \Omega}_p) &=& \frac{
  \omega_p \cos(\psi_p)\sin(\vartheta_p) }{c}, \nonumber \\
 k_{p_a,z}^{(l)}({\bf \Omega}_p) &=& \pm \frac{ n_p^{(l)}(\omega_p)\omega_p}{c}
 \cos(\vartheta_p^{(l)}), \nonumber \\
 & & \hspace{3mm} l=0,\ldots,N+1,
\label{3}
\end{eqnarray}
where the radial propagation angle $ \vartheta_p^{(l)} $ in an $ l
$-th layer obeys the Snell law:
\begin{equation}  
 n_p^{(0}\sin(\vartheta_p^{(0)}) =
 n_p^{(l)}\sin(\vartheta_p^{(l)}),
  \hspace{5mm} l=1,\ldots,N+1,
\label{4}
\end{equation}
$ \vartheta_p^{(0)} \equiv \vartheta_p $. When writing
Eq.~(\ref{3}), air around the structure was assumed ($ n_p^{(0)}
=n_p^{(N+1)} = 1 $). As the transverse components of wave vectors
do not change during the propagation, the $ x $ and $ y $
components of wave vector $ {\bf k}_p $ in Eq.~(\ref{3}) are not
indexed. Also sign $ + $ ($ - $) in Eq.~(\ref{3}) corresponds to
the forward- (backward-) propagating field.

The signal and idler fields with intensities at single-photon
level can be decomposed in the same way as the pump field in
Eq.~(\ref{2}). However, instead of coefficients $
A_{p_g,\gamma}^{(l)} $ characterizing the classical pump
amplitudes, operator coefficients $ \hat{A}_{m_a,\alpha}^{(l)} $
describing the quantized signal ($ m = s $) and idler ($ m = i $)
fields are needed \cite{Mandel1995}. The formula (\ref{2}) for the
pump field can be transformed into the form applicable to the
signal and idler fields:
\begin{eqnarray}   
 \hat{\bf E}_{m}^{(+)}({\bf r},t) &=&
  \frac{1}{\sqrt{2\pi}^3 c^2} \int_{-\pi/2}^{\pi/2}
  |\sin(\vartheta_m)| \, d\vartheta_m \int_{-\pi/2}^{\pi/2}
  d\psi_m \nonumber \\
 & & \hspace{-10mm} \int_{0}^{\infty} \omega_m^2 d\omega_m \;
   \exp \left[ i k_{m,x}({\bf \Omega_m})
  x + i k_{m,y}({\bf \Omega_m}) y\right] \nonumber \\
 & & \hspace{-10mm} \sum_{\gamma={\rm TE,TM}}  \sum_{a=F,B}
  \sum_{l=0}^{N+1} {\rm rect}^{(l)}(z) \hat{A}_{m_a,\alpha}^{(l)}({\bf
  \Omega_m}) \nonumber \\
 & & \hspace{-10mm} \mbox{} \times {\bf e}_{m,\alpha}^{(l)}({\bf \Omega_m})
  \exp\left[ ik_{m_a,z}^{(l)}({\bf \Omega}_m) (z-z_{l-1}) \right]
  \nonumber \\
 & & \hspace{-10mm} \mbox{} \times \exp(-i\omega_m t); \hspace{5mm} m=s,i.
\label{5}
\end{eqnarray}
Symbols introduced in Eq.~(\ref{5}) have the same meaning for the
signal and idler fields as those defined below Eq.~(\ref{2}) for
the pump field.

The pump electric-field amplitudes $ A_{p_F,\gamma}^{(l)} $ and $
A_{p_B,\gamma}^{(l)} $ as well as the signal and idler
electric-field operator amplitudes $ \hat{A}_{m_F,\alpha}^{(l)} $
and $ \hat{A}_{m_B,\alpha}^{(l)} $ occurring in Eqs.~(\ref{2}) and
(\ref{5}), respectively, are mutually coupled through the Fresnel
relations at the boundaries and free-space evolution inside the
layers. These relations allow to express the pump electric-field
amplitudes inside the layers in terms of the amplitudes $
A_{p_F,\gamma}^{(0)} $ and $ A_{p_B,\gamma}^{(N+1)} $
characterizing the forward- and backward-propagating incident pump
fields. On the other hand, the same relations applied to the
signal and idler fields provide the signal and idler
electric-field operator amplitudes inside the layers in terms of
operator amplitudes $ \hat{A}_{m_F,\alpha}^{(N+1)} $ and $
\hat{A}_{m_B,\alpha}^{(0)} $ that correspond to the forward- and
backward-propagating outgoing signal and idler fields. The
transfer matrix formalism describing these relations has been
developed in \cite{Yeh1988,PerinaJr2011,PerinaJr2014}. Using
quantization of photon flux \cite{Vogel2001,Huttner1990}, the
operator amplitudes $ \hat{A}_{m_F,\alpha}^{(N+1)} $ and $
\hat{A}_{m_B,\alpha}^{(0)} $ can be written using the annihilation
operators $ \hat{a}_{m_F,\alpha}^{(N+1)}({\bf \Omega}_m) $ and $
\hat{a}_{m_B,\alpha}^{(0)}({\bf \Omega}_m) $ obeying the usual
boson commutation relations:
\begin{eqnarray}  
 \hat{A}_{m_F,\alpha}^{(N+1)}({\bf \Omega}_m) &=& i \sqrt{
  \frac{\hbar\omega_m }{ 2\varepsilon_0 c }} \,
  \hat{a}_{m_F,\alpha}^{(N+1)}({\bf \Omega}_m) ; \nonumber \\
 \hat{A}_{m_B,\alpha}^{(0)}({\bf \Omega}_m) &=& i \sqrt{
  \frac{\hbar\omega_m }{ 2\varepsilon_0 c }} \,
  \hat{a}_{m_B,\alpha}^{(0)}({\bf \Omega}_m) .
\label{6}
\end{eqnarray}
Symbol $ \hbar $ stands for the reduced Planck constant. More
details can be found in \cite{PerinaJr2011,PerinaJr2014}.

An outgoing photon pair in the state $ |\psi_{s,i}^{\rm out} \rangle
$ is described by the first-order perturbation solution of the
Schr\"{o}dinger equation written as
\begin{equation}  
 |\psi_{s,i}^{\rm out}\rangle = \frac{i}{\hbar} \int_{0}^{L} dz \, \hat{G}_{\rm
 int}(z) |{\rm vac} \rangle .
\label{7}
\end{equation}
In Eq.~(\ref{7}), $ L $ denotes the structure length and $ |{\rm
vac} \rangle $ means the signal and idler vacuum state.

Substituting Eqs.~(\ref{1}), (\ref{2}), (\ref{5}), and (\ref{6})
into Eq.~(\ref{7}) we reveal the expression for the two-photon
state $ |\psi_{s,i}^{\rm out} \rangle$:
\begin{eqnarray}    
 |\psi_{s,i}^{\rm out} \rangle &=& - \frac{2i}{\sqrt{2\pi}^3 c^7}
  \sum_{l=1}^{N} \sum_{a,b,g=F,B} \sum_{\alpha, \beta, \gamma={\rm TE,TM}}
  \nonumber \\
 & & \hspace {-0.7cm} \Biggl[ \prod_{m=p,s,i}  \int_{-\pi/2}^{\pi/2} |\sin(\vartheta_m)|
  d\vartheta_m \int_{-\pi/2}^{\pi/2} d\psi_m \int_{0}^{\infty} \omega_m^2
  d\omega_m \Biggr] \nonumber \\
 & & \hspace {-0.7cm} \sqrt{ \frac{\omega_s\omega_i}{n_s^{(l)}(\omega_s)
  n_i^{(l)}(\omega_i)} } \; {\cal E}_p(\omega_p) {\cal E}_p^{\rm tr}[k_{p,x}({\bf \Omega}_p),k_{p,y}({\bf \Omega}_p)] \nonumber \\
 & & \hspace {-0.7cm} \mbox{} \times \delta(\omega_p-\omega_s-\omega_i)
  \delta \left[ k_{p,x}({\bf \Omega}_p) - k_{s,x}({\bf \Omega}_s) - k_{i,x}({\bf \Omega}_i)\right] \nonumber \\
 & & \hspace{-0.7cm} \mbox{} \times
  \delta\left[ k_{p,y}({\bf \Omega}_p) - k_{s,y}({\bf \Omega}_s) - k_{i,y}({\bf \Omega}_i) \right] \nonumber \\
 & & \hspace{-0.7cm}  \mbox{} \times \chi^{(2)(l)}({\bf \Omega}_p,{\bf \Omega}_s,{\bf \Omega}_i):
  {\bf e}_{p_g,\gamma}^{(l)}({\bf \Omega}_p) {\bf e}_{s_a,\alpha}^{(l)*}
  ({\bf \Omega}_s) {\bf e}_{i_b,\beta}^{(l)*}({\bf \Omega}_i) \nonumber \\
 & & \hspace {-0.7cm} \mbox{} \times  L_l f \left[\frac{1}{2}
  \Delta k_{g,ab,z}^{(l)}({\bf \Omega}_p,{\bf \Omega}_s,{\bf
   \Omega}_i)L_l \right] A_{p_g,\gamma}^{(l)}({\bf \Omega}_p) \nonumber \\
 & & \hspace {-0.7cm} \mbox{} \times
  \hat{a}_{s_a,\alpha}^{(l)\dagger}({\bf \Omega}_s)
  \hat{a}_{i_b,\beta}^{(l)\dagger}({\bf \Omega}_i)
  |{\rm vac} \rangle ;
\label{8}
\end{eqnarray}
$ f(x) = \exp(ix)\sin(x)/x $. Phase mismatch $ \Delta k_{g,a
b,z}^{(l)}({\bf \Omega}_p,{\bf \Omega}_s,{\bf \Omega}_i) =
k_{p_g,z}^{(l)}({\bf \Omega}_p) - k_{s_a,z}^{(l)}({\bf \Omega}_s)
- k_{i_b,z}^{(l)}({\bf \Omega}_i) $ occurs in an $ l $-th layer of
length $ L_l  = z_l-z_{l-1} $. We note that there also exist
photon pairs emitted at the boundaries
\cite{PerinaJr2009,PerinaJr2009a,PerinaJr2014} that are not
described by Eq.~(\ref{8}). Contribution of this surface SPDC
behaves similarly as the analyzed volume contribution given in
Eq.~(\ref{8}). It increases the photon-pair generation rates. We
want to point out that the second-order susceptibility $ \chi^{(2)} $ of metals
depends not only on frequencies $ \omega $ of the interacting
fields, but also on their propagation directions described by
angles $ (\theta,\psi) $ (for mode details, see Appendix~A). For
GaN layers, nonzero elements of the susceptibility tensor $
\chi^{(2)} $ take the values \cite{Miragliotta1993}
\begin{eqnarray}
 &\chi^{(2)}_{xxz}=\chi^{(2)}_{xzx}=\chi^{(2)}_{yyz}=\chi^{(2)}_{yzy}=\chi^{(2)}_{zxx}=\chi^{(2)}_{zyy}= 10\,\mbox{pm/V},&
 \nonumber \\
 &\chi^{(2)}_{zzz}=-20\,\mbox{pm/V}. & \nonumber
\end{eqnarray}

The output state $ |\psi^{\rm out}_{s,i}\rangle $ in Eq.~(\ref{8})
can be further decomposed with respect to the signal and idler
propagation directions and field polarizations. Each term describing
the signal field at position $ \mathrm{\mathbf{r}}_s $ and the idler field at position
$ \mathrm{\mathbf{r}}_i $ takes the form:
\begin{eqnarray}   
 |\psi_{s_a,i_b}^{\alpha\beta}({\bf r}_s,{\bf r}_i,t)\rangle
  &=& \prod_{m=s,i} \Biggl[ \frac{1}{c^2}
  \int_{-\pi/2}^{\pi/2}|\sin{\vartheta_m}|
   d\vartheta_m \nonumber \\
 & & \hspace{-25mm} \int_{-\pi/2}^{\pi/2} d\psi_m \int_{0}^{\infty}
   \omega_m^2 d\omega_m \Biggr] \phi_{ab}^{\alpha\beta}({\bf \Omega}_s,{\bf \Omega}_i)
   \exp[i(\omega_s+\omega_i) t]  \nonumber \\
 & & \hspace{-25mm} \mbox{} \times \exp[-i({\bf k}_{s_a}^{\rm out} {\bf r}_s + {\bf k}_{i_b}^{\rm out} {\bf r}_i)]
  \, \hat{a}_{s_a,\alpha}^\dagger({\bf \Omega}_s)
   \hat{a}_{i_b,\beta}^\dagger({\bf \Omega}_i) |{\rm vac} \rangle, \nonumber \\
 & & \hspace{0mm} a,b = F,B; \hspace{5mm}\alpha,\beta = {\rm TE, TM}.
 \label{9}
\end{eqnarray}
Wave vectors  $ {\bf k}_{s_a}^{\rm out} $ and $ {\bf k}_{i_b}^{\rm
out} $ are defined outside the structure. Spectral two-photon
amplitude $ \phi_{ab}^{\alpha\beta} ({\bf \Omega}_s,{\bf
\Omega}_i) $ defined by Eq.~(\ref{9}) gives the probability
amplitude of emitting an $ \alpha $-polarized signal photon at
frequency $ \omega_s $ and propagation direction ($ \vartheta_s,
\psi_s $) together with its $ \beta $-polarized idler twin at
frequency $ \omega_i $ and propagation direction ($ \vartheta_i,
\psi_i $) at the outputs $ a $ and $ b $ of the structure.

\section{Quantities characterizing photon pairs}

Spatial and spectral intensity properties of photon pairs
\cite{PerinaJr2006,PerinaJr2014} can be derived from the joint
signal-idler photon-number density $ n_{ab}^{\alpha\beta}({\bf
\Omega}_s,{\bf \Omega}_i) $ related to signal [idler] photons
with polarization $ \alpha $ [$ \beta $] and frequency $
\omega_s $ [$ \omega_i $] propagating at angles $
(\vartheta_s,\psi_s) $ [$(\vartheta_i,\psi_i) $] in direction $
a $ [$ b $]. Using the formula Eq.~(\ref{9}) for two-photon
state $|\psi_{s_a,i_b}^{\alpha\beta}({\bf r}_s,{\bf
r}_i,t)\rangle $ the density $ n_{ab}^{\alpha\beta} $ can be
written as follows:
\begin{equation}       
 n_{ab}^{\alpha\beta}({\bf \Omega}_s,{\bf \Omega}_i) =
  \frac{|\sin(\vartheta_s)\sin(\vartheta_i)|\omega_s^2\omega_i^2}{c^4}
  |\phi_{ab}^{\alpha\beta} ({\bf \Omega}_s,{\bf \Omega}_i)|^2 .
\label{10}
\end{equation}

Signal photon-number density $ n_{s,ab}^{\alpha\beta}({\bf
\Omega}_s) $ is then derived in the form:
\begin{eqnarray}   
 n_{s,ab}^{\alpha\beta}({\bf \Omega}_s) &=&
 \int_{-\pi/2}^{\pi/2} d\vartheta_i  \int_{-\pi/2}^{\pi/2} d\psi_i
  \int_{0}^{\infty} d\omega_i \, n_{ab}^{\alpha\beta}({\bf \Omega}_s,{\bf \Omega}_i).
  \nonumber \\
 & &
\label{11}
\end{eqnarray}
Subsequently, the signal spectral photon-number density $
n_{s,ab}^{\omega, \alpha\beta}(\Omega_s) $ is determined along the
formula:
\begin{eqnarray}   
 n_{s,ab}^{\omega,\alpha\beta}(\omega_s) = \int_{-\pi/2}^{\pi/2} d\vartheta_s
 \int_{-\pi/2}^{\pi/2} d\psi_s
 \, n_{s,ab}^{\alpha\beta}({\bf \Omega}_s).
\label{12}
\end{eqnarray}
Similarly, the signal transverse photon-number density $
n_{s,ab}^{{\rm tr},\alpha\beta}(\vartheta_s,\psi_s) $
characterizing photons propagating in direction ($
\vartheta_s,\psi_s $) is determined as:
\begin{eqnarray}   
 n_{s,ab}^{{\rm tr},\alpha\beta}(\vartheta_s,\psi_s) = \int_{0}^{\infty}
  d\omega_s \, n_{s,ab}^{\alpha\beta}({\bf \Omega}_s).
\label{13}
\end{eqnarray}

Intensity correlations between the signal and idler fields in
their transverse planes are described by the joint signal-idler
transverse photon-number density $ n_{ab}^{{\rm
cor},\alpha\beta}(\vartheta_s,\psi_s,\vartheta_i,\psi_i ) $
characterizing a photon pair with signal [idler] photon
propagating along angles ($ \vartheta_s,\psi_s $) [($
\vartheta_i,\psi_i $)] in direction $ a $ [$ b $]:
\begin{eqnarray}   
 n_{ab}^{{\rm cor},\alpha\beta}(\vartheta_s,\psi_s,\vartheta_i,\psi_i)
  &=& \int_{0}^{\infty} d\omega_s \int_{0}^{\infty} d\omega_i \,
  n_{ab}^{\alpha\beta}({\bf \Omega}_s,{\bf \Omega}_i). \nonumber\\
 & &
\label{14}
\end{eqnarray}
If a signal photon is detected at angle ($ \vartheta_s^0,\psi_s^0
$), the joint signal-idler transverse photon-number density $
n_{ab}^{{\rm
cor},\alpha\beta}(\vartheta_s^0,\psi_s^0,\vartheta_i,\psi_i )$
gives the probability of detecting the accompanying idler photon
at direction $(\vartheta_i,\psi_i )$. This probability determines
the shape of correlated area \cite{Hamar2010}.

In the time domain, two-photon states are characterized by a
two-photon temporal amplitude $ {\cal A}(\tau_s,\tau_i) $ that
gives the probability amplitude of detecting a signal photon at
time $ \tau_s $ together with detecting the accompanying idler
photon at time $ \tau_i $. Using two-photon spectral amplitude $
\phi_{ab}^{\alpha\beta} $ in Eq.~(\ref{9}), the two-photon
temporal amplitude $ {\cal A}(\tau_s,\tau_i) $ can be expressed
as:
\begin{eqnarray} 
 {\cal A}_{ab}^{\alpha\beta}(\theta_s,\psi_s,\tau_s,\theta_i,\psi_i,\tau_i)
  &=& \frac{\sqrt{|\sin(\vartheta_s)\sin(\vartheta_i)|}\hbar }{
  4\pi \varepsilon_0 c^3}\int_{-\infty}^{\infty} d\omega_s \nonumber\\
 & & \hspace{-38mm} \int_{-\infty}^{\infty} d\omega_i
  \sqrt{\omega_{s}^3\omega_{i}^3}
  \phi_{ab}^{\alpha\beta}({\bf \Omega}_s,{\bf \Omega}_i)
  \exp(-i\omega_s\tau_s) \exp(-i\omega_i\tau_i). \nonumber \\
 & &
\label{15}
\end{eqnarray}

Temporal properties of photon pairs are usually experimentally
investigated employing the Hong-Ou-Mandel interferometer
\cite{Hong1987}. In this interferometer, two photons are mutually
delayed by $\tau_l$ and then they interfere on a beam splitter
which output ports are monitored by two detectors measuring in
coincidence. A normalized coincidence-count rate $ R $ depends on
time delay $ \tau_l $ according to the formula:
\begin{equation} 
 R_{ab}^{\alpha\beta}(\tau_l,\vartheta_s,\psi_s,\vartheta_i,\psi_i)=
  1 -\rho_{ab}^{\alpha\beta}(\tau_l,\vartheta_s,\psi_s,\vartheta_i,\psi_i),
 \label{16}
\end{equation}
where
\begin{eqnarray} 
 \rho_{ab}^{\alpha\beta}(\tau_l,\vartheta_s,\psi_s,\vartheta_i,\psi_i) &=&
  \frac{|\sin(\vartheta_s)\sin(\vartheta_i)|\hbar^2 }{ 2 c^4 R_{0,ab}^{\alpha\beta}}
  \int_{0}^{\infty} d\omega_s \nonumber \\
 & & \hspace{-35mm} \int_{0}^{\infty} d\omega_i \omega_s^3\omega_i^3
  {\rm Re} \Bigl\{ \phi_{ab}^{\alpha\beta *}({\bf \Omega}_s,{\bf \Omega}_i)
  \phi_{ab}^{\alpha\beta}(\omega_i,\vartheta_s,\psi_s,\omega_s,\vartheta_i,\psi_i) \nonumber \\
 & & \hspace{-20mm} \times
  \exp[i(\omega_s-\omega_i)\tau_l] \Bigr\} ,
\label{17} \\
 R_{0,ab}^{\alpha\beta}(\vartheta_s,\psi_s,\vartheta_i,\psi_i) &=&
  \frac{|\sin(\vartheta_s)\sin(\vartheta_i)|\hbar^2 }{2c^4} \int_{0}^{\infty} d\omega_s
  \nonumber \\
 & & \hspace{-20mm} \int_{0}^{\infty} d\omega_i
  \, \omega_s^3\omega_i^3 |\phi_{ab}^{\alpha\beta}({\bf \Omega}_s,{\bf \Omega}_i)|^2 . \nonumber
\end{eqnarray}

Enhancement of the nonlinear interaction inside a layered
structure originates from increased electric-field amplitudes due
to back-scattering on the boundaries. This enhancement can be
quantified using a reference structure defined in
\cite{PerinaJr2011}. This reference structure uses the natural
material nonlinearity exploiting the greatest nonlinear
coefficient, but it does not contain any boundary that would
scatter the propagating light. The reference structure generates a
signal photon in direction ($ \vartheta_s,\psi_s $) together with
an idler photon in direction ($ \vartheta_i,\psi_i $) exploiting
phase matching in the transverse plane reached with a pump plane
wave found in the spatial spectrum $ {\cal E}_p^{\rm tr} $. The
corresponding two-photon state $ |\psi_{s,i}^{\rm ref}\rangle $ is
expressed as:
\begin{eqnarray}    
 |\psi_{s,i}^{\rm ref}\rangle &=&
  - \frac{2i}{\sqrt{2\pi}^3 c^5 } \Biggl[ \prod_{m=s,i}
  \int_{-\pi/2}^{\pi/2} |\sin(\vartheta_m)| \, d\vartheta_m
  \nonumber \\
 & & \int_{-\pi/2}^{\pi/2}
  d\psi_m \int_{0}^{\infty} \, \omega_m^2 d\omega_m \Biggr]
  {\cal E}_{p}(\omega_s+\omega_i) \nonumber \\
 & & \mbox{} \times
  {\cal E}_p^{\rm tr} \left[k_{s,x}({\bf \Omega}_s) + k_{i,x}({\bf
  \Omega}_i),k_{s,y}({\bf \Omega}_s) + k_{i,y}({\bf \Omega}_i)
  \right] \nonumber \\
 & & \mbox{} \times \sum_{l=1}^{N}
  \sqrt{ \frac{\omega_s \omega_i}{n_s^{(l)}(\omega_s) n_i^{(l)}(\omega_i)} }  \,
  \max(|\chi^{(2)(l)}|)\, L_l \,  \nonumber \\
 & & \mbox{} \times  \hat{a}_{s}^\dagger({\bf \Omega}_s) \hat{a}_{i}^\dagger({\bf \Omega}_i)
  |{\rm vac} \rangle.
\label{18}
\end{eqnarray}
Creation operator $ \hat{a}_{s}^{\dagger}({\bf \Omega}_s) $ [$
\hat{a}_{i}^{\dagger}({\bf \Omega}_i) $] describes the signal
[idler] photon at the output plane of the structure. Function $
\max $ gives the maximal value of elements of nonlinear tensor $
\chi^{(2)(l)} $. Using the signal photon-number density $n_s^{\rm
ref}({\bf \Omega}_s) $ of the reference structure given in
Eq.~(\ref{11}), the signal relative photon-number density $
\eta_{s,ab}^{\alpha\beta}({\bf \Omega}_s) $ at frequency $
\omega_s $ and in emission direction ($ \vartheta_s,\psi_s $) is
conveniently defined using the relation:
\begin{equation}   
 \eta_{s,ab}^{\alpha\beta}({\bf \Omega}_s) = \frac{
  n_{s,ab}^{\alpha\beta}({\bf \Omega}_s) }{ \max_{\vartheta_s,\omega_s}\left[n_s^{\rm ref}({\bf
  \Omega}_s)\right]  }.
\label{19}
\end{equation}
In Eq.~(\ref{19}), the maximum is taken over the whole interval of
radial emission angles $ \vartheta_s $ and frequencies $ \omega_s
$ assuming a fixed azimuthal emission angle $ \psi_s^0 $.

In our numerical calculations, we consider a cw pump field with
amplitude $ \xi_p $ and Gaussian transverse profile, i.e.
\begin{eqnarray}  
 {\cal E}_p(\omega_p) &=& \xi_p \delta (\omega_p-\omega_p^0) ,
\label{20} \\
 {\cal E}_p^{\rm tr}(k_x,k_y) &=& \frac{r_p}{\sqrt{2\pi}}
  \exp\left[ - \frac{ r^2_p(k_x^2 + k_y^2)}{4} \right] ;
\label{21}
\end{eqnarray}
$ \omega_p^0 $ is the central frequency and $ r_p $ stands for the
radius of transverse profile. It holds that $ \int dk_x \int dk_y
|{\cal E}_p^{\rm tr}(k_x,k_y)|^2 = 1 $. Whenever the
expression $ \delta^2(\omega) $ occurs in the above defined
formulas, it has to be replaced by the expression $ 2T/(2\pi)
\delta(\omega)$ obtained for the fields defined inside interval $
(-T,T) $. Physical quantities obtained per unit time interval are
reached in the limit $ T \rightarrow \infty $.

\section{A simple metal-dielectric resonator}

Though both the metal and dielectric layers are nonlinear, the
dielectric layers are able to provide much higher photon-pair
fluxes. For this reason, the presence of thin metal layers is
important for an enhancement of electric-field amplitudes inside
the structure. This enhancement then results in much stronger
nonlinear interaction and efficient production of photon pairs.
Compared to pure dielectric layered structures like those composed
of GaN and AlN, analyzed in \cite{PerinaJr2011,PerinaJr2014},
metal-dielectric layered structures allow for much higher
enhancement of electric-field amplitudes due to the high
refraction-index contrast of the used metal and dielectric
materials. For comparison and considering the wavelength 800~nm,
this contrast equals 2.51 [2.16] for GaN [AlN] layers and 5.3 [2.51] for
Ag [GaN] layers analyzed here. However, strong attenuation and
losses of the electric-field amplitudes occur in metal layers.
This puts restrictions to the possible thicknesses of metal layers
as well as to the number of metal layers embedded into the
structure.

To get deeper insight into the behavior of metal-dielectric
layered structures, we first consider the simplest possible
structure composed of only one nonlinear GaN layer sandwiched by
two thin Ag layers. Thus, the Ag layers form mirrors of a simple
resonator that enhances the electric-field amplitudes inside the
GaN layer. To achieve efficient nonlinear interaction, we apply
the method for designing an efficient layered structure for SPDC
suggested in \cite{PerinaJr2011}. Lengths $ l_2 $ of GaN layers
and $ l_1 $ of Ag layer vary in the method to reveal the most
efficient structure. In the method, only pairs $ (l_1,l_2) $ of
lengths that provide transmission maxima for the pump field at a
chosen wavelength $ \lambda_p^0 $ are analyzed. Concentrating on
the highest transmission maximum that also gives the greatest
enhancement of the pump field, the appropriate pairs $ (l_1,l_2) $
of lengths form a one-dimensional parametric system. This means
that for any value of GaN-layer length $ l_2 $ there exists only
one value of Ag-layers length $ l_1 $.

\begin{figure} 
 \includegraphics[scale=1]{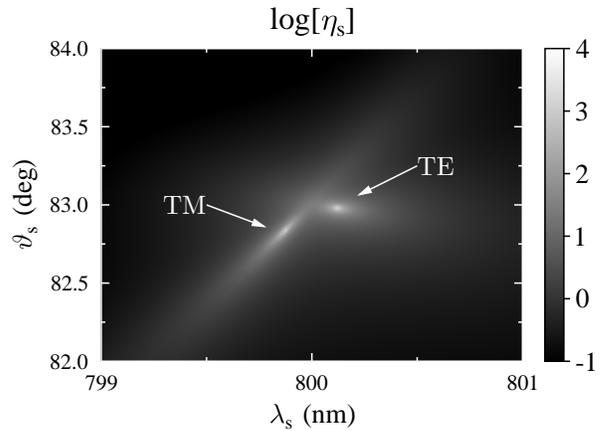}
 \caption{Topo graph of relative signal photon-number density
 $ \eta_s $ in dependence on signal radial emission
 angle $\vartheta_s$ and wavelength $\lambda_s$ for a simple
 'metal-dielectric' resonator structure composed of one GaN layer and
 two Ag layers. Both photons with arbitrary polarizations propagate
 along the $ +z $ axis; $  \lambda_p^0 = 400 $~nm,
 $l_1=95.1195$~nm, $l_2=9.6376$~nm, $\psi_s^0=0$~deg;
 log denotes the decimal logarithm.}
 \label{fig2}
\end{figure}

In the analysis, we consider a plane-wave TE-polarized pump-field
at central wavelength $\lambda_p^0=400$~nm impinging on the
structure at normal incidence. Structures with thick Ag layers
($l_1 > 10$~nm) provide frequency-degenerated photon pairs. On the
other hand, structures with thin Ag layers emit frequency
non-degenerated photon pairs. The greatest value of relative
signal photon-number density $ \eta $ defined in Eq.~(\ref{19}) is
reached for slightly frequency non-degenerated photon-pair
emission for $l_1=9.6376$~nm and $l_2=95.1195$~nm. We note that
the signal and idler photons can leave the structure either along
the $ +z $ or $ -z $ axes, so four possible combinations for
photon pairs exist. Nevertheless different photon pairs have
comparable properties. That is why, we pay attention to only
photon pairs with both photons propagating along the $ +z$
direction. The structure generates photon pairs around the radial
emission angle $\vartheta =83$~deg. Two emission maxima in
relative signal photon-number density $ \eta_s $ plotted in
Fig.~\ref{fig2} are observed. Whereas one maximum contains
TE-polarized photons, the other maximum is composed of
TM-polarized photons. As elements $\chi^{(2)}_{xxz} $ and $
\chi^{(2)}_{xzx}$ of susceptibility tensor participate in the
nonlinear interaction, a TE-polarized photon is accompanied by a
TM-polarized photon and vice versa. Two maxima in relative signal
photon-number density $ \eta_s $, shown in Fig.~\ref{fig2}, are
sharp compared to similar dielectric structures. This is a
consequence of strong interference of back-scattered optical
fields caused by the high refractive-index contrast. These sharp
features are characteristic for both spectral and spatial
properties of photon pairs.

\begin{figure} 
\includegraphics[scale=1]{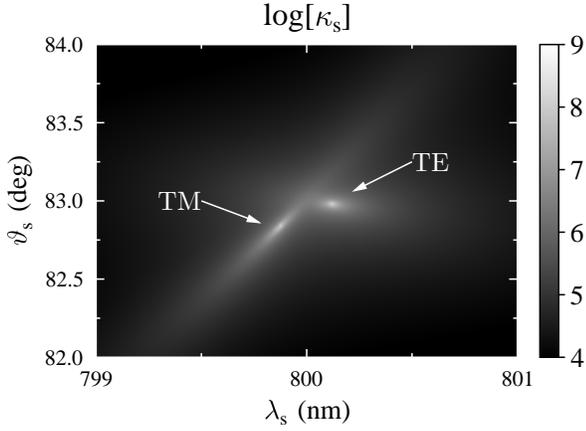}
\caption{Topo graph of ratio $\kappa $ of signal photon-number
densities $ n_s $ of the simple 'metal-dielectric' resonator
structure and GaN monolayer of equal thickness as it depends on
signal radial emission angle $\vartheta_s$ and wavelength $
\lambda_s$. Parameters are written in the caption of
Fig.~\ref{fig2}.}
\label{fig3}
\end{figure}

The advantage of 'metal resonator' surrounding the nonlinear GaN
layer can be quantified comparing its signal photon-number density
$ n_s $ [Eq.~(\ref{11})] with that characterizing one GaN
monolayer structure of the same length ($l=114.3947$~nm). Ratio $
\kappa $ of these photon-number densities $ n_s $ (see
Fig.~\ref{fig3}) shows that the enhancement of up to five orders
in magnitude is reached in areas of maximal emission intensities,
i.e. under conditions of the strongest constructive interference
of the signal [idler] field. The enhancement factor rapidly drops
down when wavelengths $ \lambda_s $ and radial emission angles $
\vartheta_s $ move away from these optimal conditions.

\section{An efficient metal-dielectric structure}

In order to sufficiently enhance the nonlinear interaction, more
complex metal-dielectric layered structures have to be considered.
There exists an interval of suitable numbers of the used layers.
On one side, larger number of layers leads to strong interference
and also to high enhancement of electric-field amplitudes. On the
other side, larger number of metal layers results in strong
attenuation of the electric fields. To keep balance between these
effects, we have decided to design a structure with five metal Ag
layers sandwiched by six GaN layers (for the scheme, see
Fig.~\ref{fig1}).

\begin{figure} 
 \centering \subfigure[]{\includegraphics[scale=1]{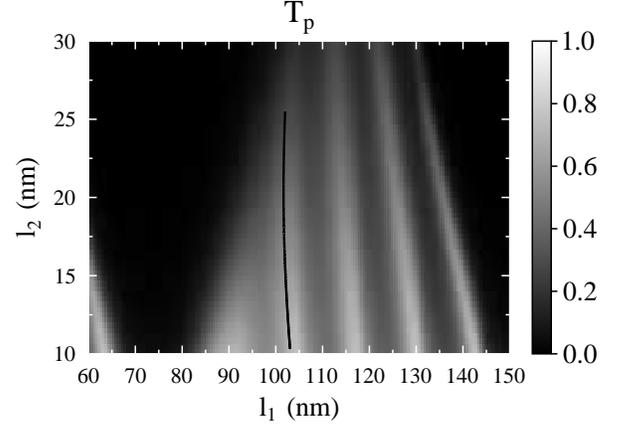}}
 \subfigure[]{\includegraphics[scale=1]{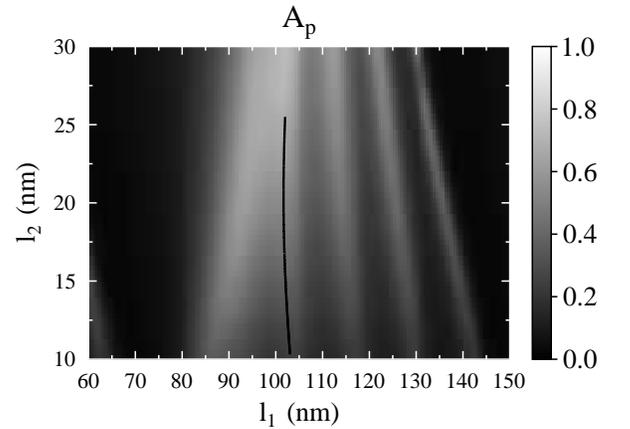}}
 \caption{Topo graphs of (a) intensity transmission coefficient $
 T_p $ and (b) intensity absorption coefficient $ A_p $ depending on
 layers' lengths $l_1$ and $l_2$ for TE-polarized field at $
 \lambda_p^0 = 400 $~nm. Positions of maxima in the first
 transmission band are indicated by solid black curves.}
\label{fig4}
\end{figure}

Following the design procedure, we have plotted the pump-field
intensity transmission coefficient $ T_p $ at the wavelength $
\lambda_p^0 = 400 $~nm and for TE polarization [see
Fig.~\ref{fig4}(a)] as it depends on layers' lengths $ l_1 $ and $
l_2 $. The pump field impinging on the structure at normal
incidence has been assumed. In this graph, five transmission bands
can be seen. It follows from the theory of band-gap structures
that the greatest enhancement of electric-field amplitudes occurs
in the transmission band closest to the band gap. In this band,
also the greatest values of absorption $ A_p $ are found [see
Fig.~\ref{fig4}(b)] indicating large electric-field amplitudes
inside the metal layers \cite{Larciprete2008}.

\begin{figure} 
 \centering \includegraphics[scale=1]{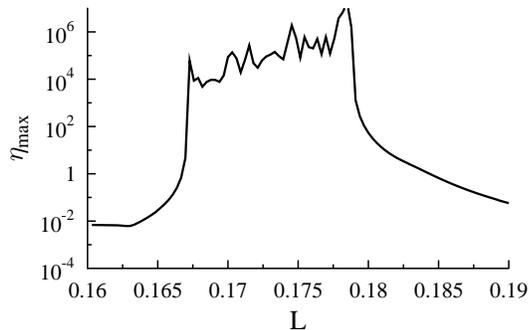}
 \caption{Maximum $\eta_s^{\rm max} $ of relative signal
 photon-number density $ \eta_s $ depending on ratio $ L $ of layers'
 lengths, $ L=l_2/l_1$, for structures composed of 11 layers such that the
 pump field at $ \lambda_p^0 = 400 $~nm occurs in the center of the first
 transmission band (see Fig.~\ref{fig4}).}
\label{fig5}
\end{figure}

Structures corresponding to the maxima of the first transmission
band have been parameterized by the ratio $L=l_2/l_1$. Maximum $
\eta_s^{\rm max} $ of relative signal photon-number density $
\eta_s $ taken over frequency $ \omega_s $ and radial emission
angle $ \vartheta_s $ assuming fixed azimuthal angle $\psi_{s,0}$
was chosen for quantification of efficiency of the nonlinear
process. Structures with parameter $ L $ in the interval $
(0.1,0.25)$ were only considered because very thin metal layers do
not sufficiently enhance the electric-field amplitudes. Moreover,
their transmission bands are broader. On the other hand, thick
metal layers attenuate the propagating electric fields. Maximal
values $ \eta_s^{\rm max} $ of relative signal photon-number
density $ \eta_s $ were found in two regions: $L\in(0.17,0.18)$
and $L \in (0.225,0.24)$. In these regions, $\eta_s^{\rm max} $
reaches values around $ 10^6 $. The first region of $ L $ analyzed
in Fig.~\ref{fig5} is more suitable and contains the most
efficient structure ($L=0.178$) with lengths $l_1=101.752$~nm and
$l_2=18.083$~nm. The obtained values of maxima $ \eta_s^{\rm max}
$ are higher by two orders in magnitude compared to the values of
maxima $ \eta_s^{\rm max} $ of the 'metal resonator' investigated
in Sec.~IV. Additionally, these values are even higher by seven
orders in magnitude compared to those of pure dielectric layered
structures studied in \cite{PerinaJr2011}. Detailed analysis of
SPDC inside the metal-dielectric structures shows that dielectric
layers are the major source of photon pairs. Metal layers give
photon-pair numbers lower by six orders in magnitude compared to
the dielectric layers. Nevertheless, they play a critical role in
the enhancement of electric-field amplitudes inside the structure
due to their high indices of refraction. We have also analyzed
SPDC involving a TM-polarized pump field along the same vein.
However, the obtained values of maxima $ \eta_s^{\rm max} $ have
been found considerably lower than those discussed above for the
TE-polarized pump field.

\begin{figure} 
 \centering \subfigure[]{\includegraphics[scale=1]{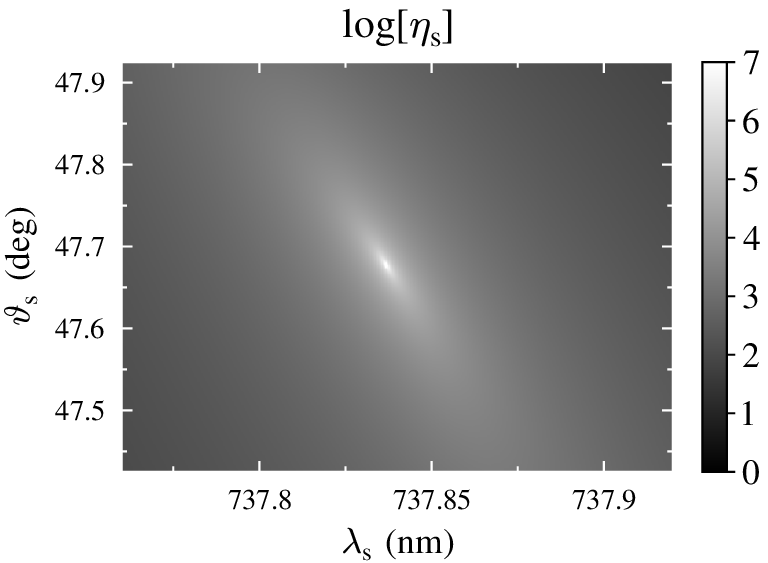}}
 \subfigure[]{\includegraphics[scale=1]{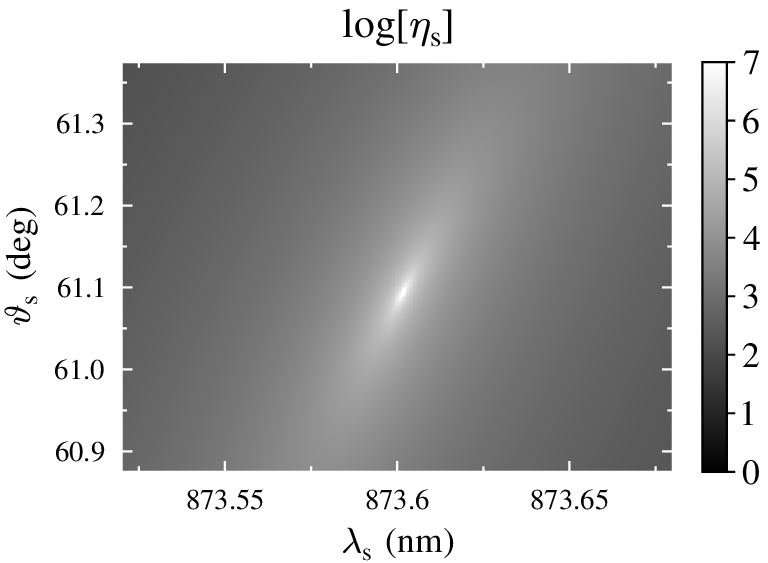}}
 \caption{Topo graphs of relative signal photon-number density $
 \eta_s $ in dependence on signal wavelength $ \lambda_s $
 and radial emission angle $\vartheta_s $ for two regions
 containing (a) TE-polarized and (b) TM-polarized photons;
 $ \lambda_p^0 = 400 $~nm, $l_1=101.752$~nm, $l_2=18.083$~nm.}
\label{fig6}
\end{figure}

Relative signal photon-number density $ \eta_s $ of this structure
(plotted in Fig.~\ref{fig6}) reveals two emission peaks. One peak
is centered at the wavelength $\lambda_s=737.837$~nm and the
radial emission angle $ \vartheta_s=47.686 $~deg, the other peak
occurs at the wavelength $\lambda_s=873.601$~nm and the radial
emission angle $\vartheta_s=61.095$~deg. The signal photon at
wavelength $\lambda_s=737.837$~nm is TE polarized and its twin has
TM polarization. On the other hand, the signal photon at
wavelength $\lambda_s=873.601$~nm has TM polarization, whereas its
twin is TE polarized. This means that the first photon pair
exploits the element $ \chi^{(2)}_{xxz} $ of susceptibility tensor
whereas the second photon pair uses the element $ \chi^{(2)}_{xzx}
$. The emission peaks are very narrow in both the wavelength $
\lambda_s $ and radial emission angle $\theta_s$. The intensity
peaks' widths $ \Delta\lambda_s $ are narrower than $1\times
10^{-3}$~nm (full width at half maximum, FWHM). In radial emission
angle, the intensity peaks' widths $ \Delta \theta_s $ are
narrower than $5 \times 10^{-2}$~deg. It is worth to stress that
the sharpness of these peaks arises from the behavior of
TM-polarized fields. The analyzed system has nearly radial
symmetry which is only weakly broken by the varying values of $
\chi^{(2)} $ elements in azimuthal direction. So the emitted
photon pairs form two narrow concentric rings; slightly changing
intensities are found around these rings.

\begin{figure} 
 \begin{center}
 \subfigure[]{\includegraphics[scale=1]{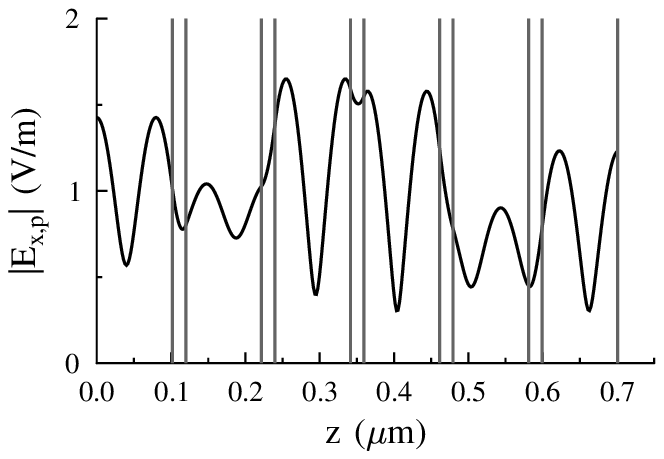}}
  \subfigure[]{\includegraphics[scale=1]{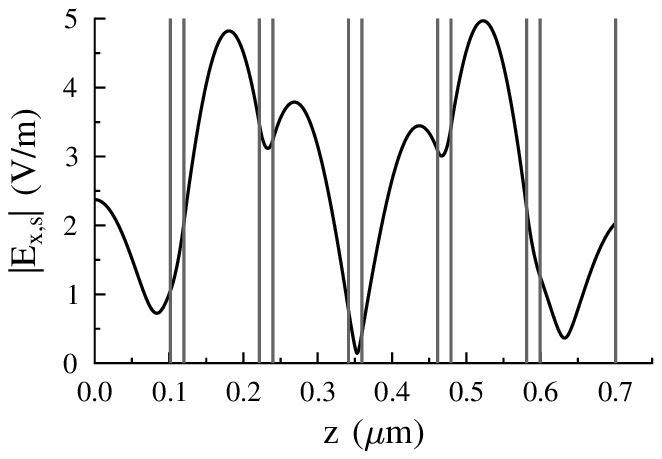}}
 \subfigure[]{\includegraphics[scale=1]{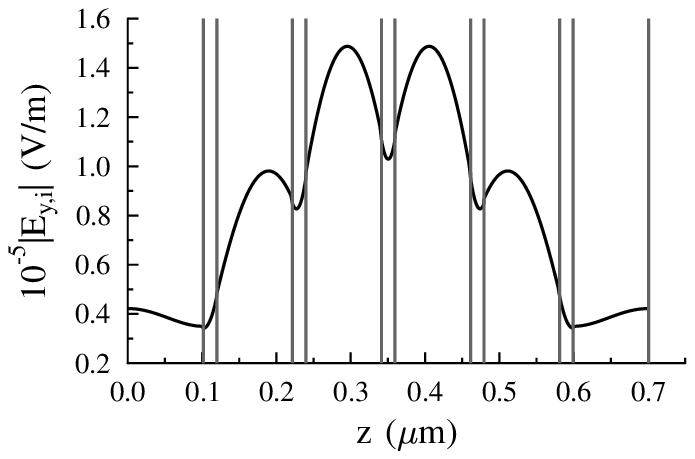}}
 \end{center}
 \caption{Profile of modulus of the electric-field amplitude for
  (a) pump, (b) signal and (c) idler fields along the $ z $ axis for
  the pump field with amplitude 1~V/m incident at $ z = 0$~m and
  outgoing signal and idler fields with amplitudes 1~V/m at the
  end of the structure composed of eleven GaN/Ag layers described in the
  caption to Fig.~\ref{fig6}. In the TM-polarized idler field, the $ z $
  component of electric-field amplitude is by several orders in
  magnitude lower than the plotted $ y $ component;
  $ \lambda_p = 400 $~nm, $ \vartheta_p = 0 $~deg,
  $ \lambda_s=737.8367$~nm, $\vartheta_s=47.686$~deg,
  $ \lambda_i=873.6015$~nm, $\vartheta_i=-61.095$~deg.}
\label{fig7}
\end{figure}

The electric-field amplitude profiles of the interacting fields
along the propagating $ z $ axis for $ (p,s,i) = {\rm (TE,TE,TM)}
$ interaction are shown in Fig.~\ref{fig7}. The pump
electric-field amplitude profile is determined for the incident
electric-field amplitude 1~V/m impinging on the structure at $ z =
0 $~m. The signal and idler electric-field amplitude profiles are
such that they give the outgoing amplitude 1~V/m at the end of the
structure and 0~V/m for the outgoing amplitude at $ z = 0 $~m. The
TE-polarized pump and signal fields have their electric-field
amplitudes inside the structure enhanced several times. In
contrast, the enhancement factor of TM-polarized idler field
equals around $ 10^5 $ due to highly constructive interference of
the back-scattered fields at the boundaries. For comparison, the
enhancement factor for GaN/AlN layered structures typically equals
several tens \cite{PerinaJr2011}.

\begin{figure} 
 \subfigure[]{\includegraphics[scale=1]{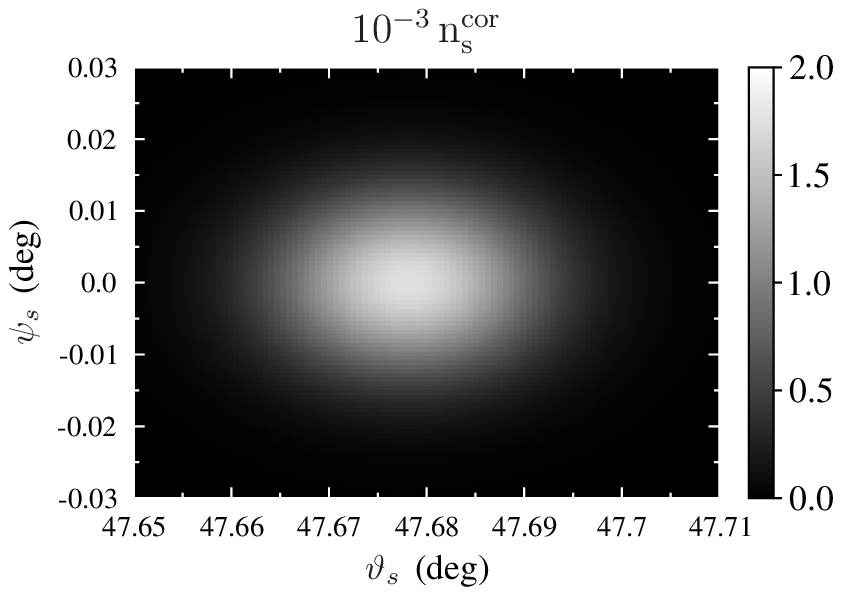}}
 \subfigure[]{\includegraphics[scale=1]{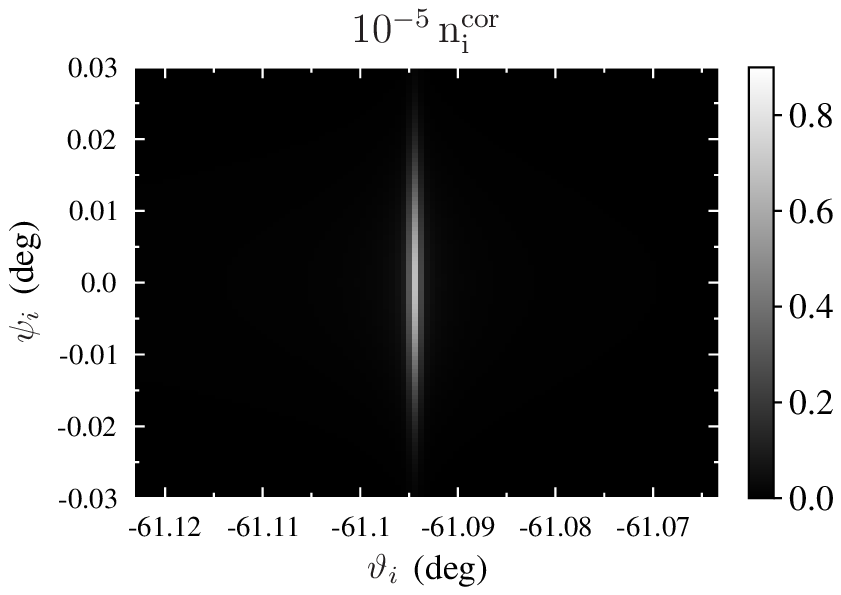}}
 \caption{Correlated area $ n^{\rm cor} $ of (a) signal [(b) idler]
 photon observed
 after detection of an idler [signal] photon at direction $
 \vartheta_i^0=-61.095
 $~deg and  $\psi_i^0=0$~deg [$ \vartheta_s^0= 47.686$~deg
 and
 $\psi_s^0=0$~deg] for the structure analyzed in Fig.~\ref{fig6}.
 The correlated
 areas are  normalized such that $\int d\vartheta \int d\psi \, n^{\rm
 cor}(\vartheta,\psi) = (\pi/180)^2$.}
\label{fig8}
\end{figure}

Also correlated areas characterizing spatial correlations between
the signal and idler intensities are narrow. Two different shapes
of correlated areas found in the analyzed structure are shown in
Fig.~\ref{fig8} for a pump beam with Gaussian transverse profile
of radius $r_p=1$~mm. If we fix the emission direction of the
TM-polarized idler photon at $ \vartheta_i=-61.095$~deg, the
correlated area of TE-polarized signal photon has roughly a
Gaussian shape which originates in the Gaussian pump-field
transverse shape [see Fig.~\ref{fig8}(a)]. On the other hand, when
the TE-polarized signal photon is detected at
$\vartheta_s=47.686$~deg, the correlated area of TM-polarized
idler photon is highly elliptic [see Fig.~\ref{fig8}(b)]. The
reason is that its extension along the azimuthal angle $ \psi_i $
is determined by the pump-beam radius $ r_p $, whereas its
extension along the radial angle $ \vartheta_i $ is strongly
limited by the properties of TM modes related to their strong
back-scattering on the boundaries. The dependence on pump-beam
radius $ r_p $ can be used to tailor the extensions of correlated
areas \cite{Hamar2010}.

\section{Temporal properties of emitted photon pairs}

\begin{figure} 
 \subfigure[]{\includegraphics[scale=1]{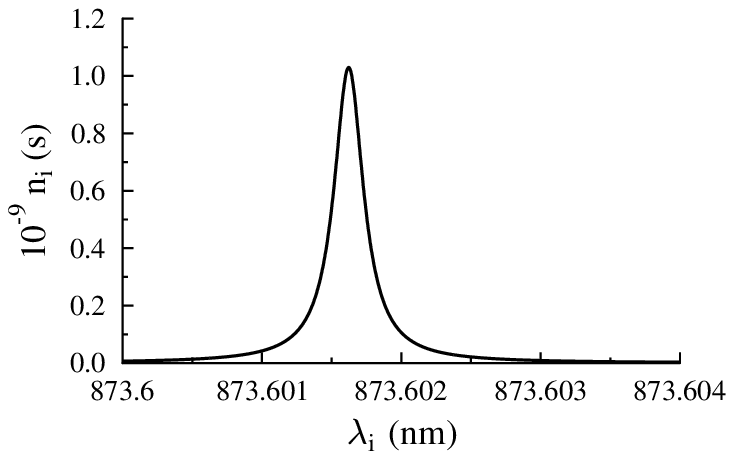}}
 \subfigure[]{\includegraphics[scale=1]{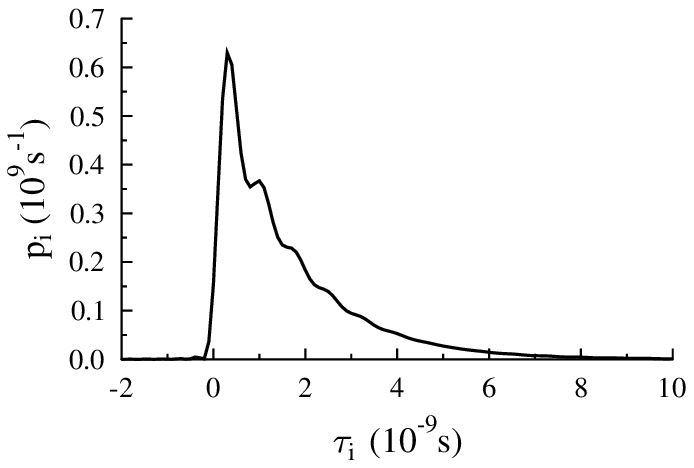}}
 \caption{(a) Idler spectral photon-number density $ n_i $ as a
 function of idler wavelength $ \lambda_i $ and (b) probability
 density $ p_i $ of detecting an idler photon at time $ \tau_i
 $ provided that its signal twin was detected at time $ \tau_s
 = 0 $~s; $ p_i(\tau_i) = C|{\cal A}(\tau_s=0,\tau_i)|^2 $ using
 an appropriate normalization constant $ C $. A photon pair is
 emitted in directions $\vartheta_s=47.686$~deg and $
 \psi_s=0$~deg and $\vartheta_i=-61.095$~deg and $ \psi_i=0$~deg
 in the structure described in the caption to Fig.~\ref{fig6}.
 Normalization is such that $ \int
 d\omega_i \, n_i(\omega_i) = 1 $ and $ \int d\tau_i \,
 p_i(\tau_i) = 1 $.}
\label{fig9}
\end{figure}

Due to stationarity, the two-photon spectral amplitude $
\phi(\omega_s,\omega_i) $ gets a general form $ f_i(\omega_i)
\delta(\omega_p^0 -\omega_s-\omega_i) $, in which the $ \delta
$-function expresses the energy conservation law. The squared
modulus $ |f_i|^2 $ is then linearly proportional to the idler
spectral photon-number density $ n_i^\omega(\omega_i) $. For the
analyzed structure, the spectral density $ n_i^\omega $ of a
photon pair with signal photon propagating along direction $
\vartheta_s^0= 47.686$~deg and $\psi_s^0=0$~deg and idler photon
propagating along direction $ \vartheta_i^0=-61.095 $~deg and
$\psi_i^0=0$~deg attains the form of a very narrow peak of width $
4.45\times 10^{-4}$~nm [FWHM, see Fig.~\ref{fig9}(a)].

The narrow spectral peak is responsible for longer temporal
correlations of fields' intensities compared to those
characterizing photon pairs generated in a typical bulk crystals.
For the analyzed structure and cw pumping, intensity temporal
correlations occur at the time scale of ns [for the conditional
probability density $ p_i $ of detecting an idler photon at time $
\tau_i $, see Fig.~\ref{fig9}(b)]. It is worth noting that the
signal- and idler-field group velocities considerably differ. The
TE-polarized signal photons propagate on average faster than the
TM-polarized idler photons that undergo on average much higher
number of back reflections on the boundaries after their emission.
If pulsed SPDC occurred in the structure, the idler-field
detection interval would be much wider than that of the signal
field.

\begin{figure} 
 \centering
  \includegraphics[scale=1]{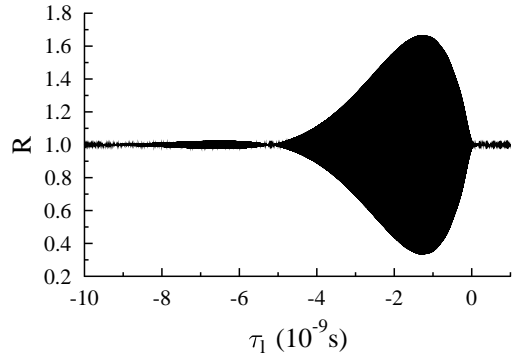}
  \caption{Normalized coincidence count rate $R$ in the
  Hong-Ou-Mandel interferometer depending on mutual time
  delay $ \tau_l $ between the signal and idler photons.
  The structure described in the caption to Fig.~\ref{fig6} is analyzed.}
  \label{fig10}
 \end{figure}

Different group velocities of the signal and idler photons inside
the structure also result in highly asymmetric coincidence-count
rate profiles observed in the Hong-Ou-Mandel interferometer, as
documented in Fig.~\ref{fig10}. In this interferometer, a much
longer average delay of the idler photon has to be compensated by
a delay line placed into the signal-photon path to achieve mutual
interference of both photons at a beam splitter. Fast oscillations
caused by nonzero difference of the signal and idler central
frequencies are also visible in the normalized coincidence-count
rate $ R $ in Fig.~\ref{fig10}. We note that the Hong-Ou-Mandel
interferometer represents the simplest tool for the observation of
temporal correlations between photons.

\section{Losses in the structure and noise photons}

\begin{figure} 
 \centering \subfigure[]{\includegraphics[scale=1]{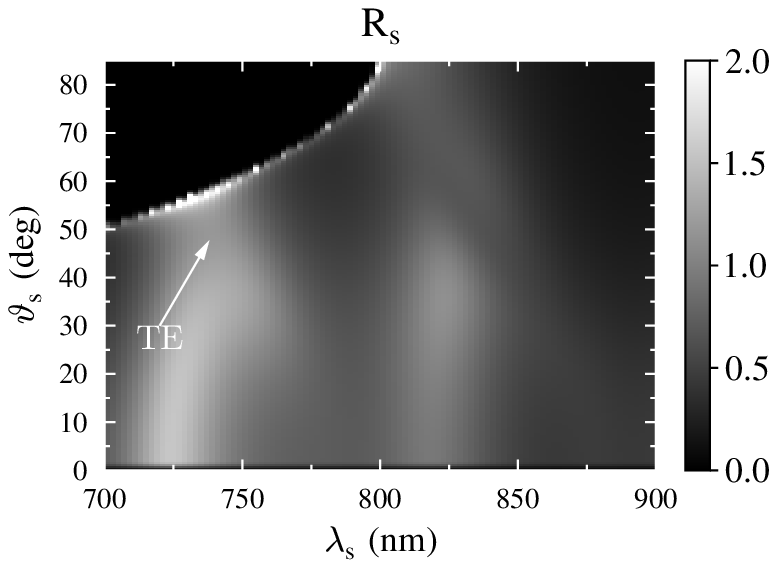}}
 \subfigure[]{\includegraphics[scale=1]{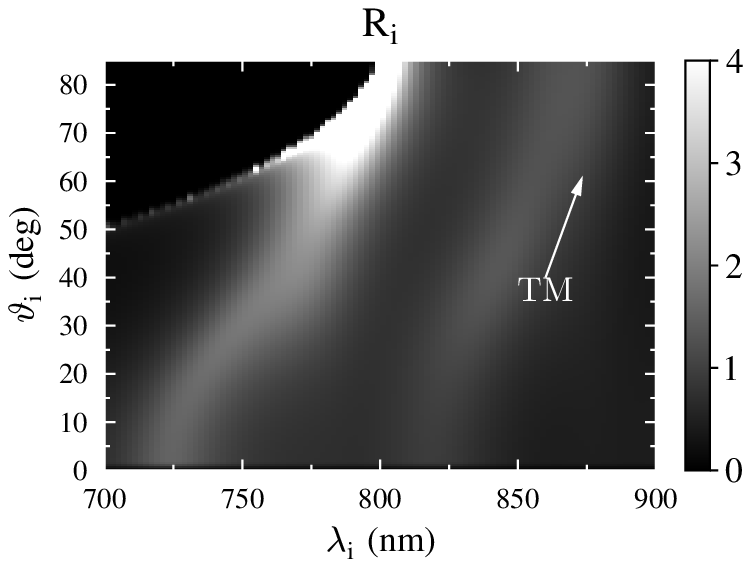}}
 \caption{Topo graph of ratio (a) $ R_{s,FF}^{\rm TE,TM} $
  [(b) $ R_{i,FF}^{\rm TE,TM} $] of signal [idler] noise
  photon-number density and photon-pair density
  in dependence on signal [idler] radial emission angle
  $ \vartheta_s $ [$ \vartheta_i $] and wavelength $ \lambda_s $
  [$ \lambda_i $] determined along Eq.~(\ref{B5}) in Appendix B.
  The photon-pair field contains the forward-propagating $ {\rm TE}
  $-polarized signal and $ {\rm TM} $-polarized idler photons;
  $ \lambda_p^0 = 400 $~nm, $l_1=101.752$~nm, $l_2=18.083$~nm.}
\label{fig11}
\end{figure}

Non-negligible losses occur in the analyzed metal-dielectric
layered structures because of the presence of highly absorbing
metal layers. When one photon from a photon pair is absorbed
whereas the other photon leaves the structure, the emitted joint
signal and idler field contains also the single-photon noise
present both in the signal and idler fields. According to the
theory developed in Appendix~B, these noise contributions are
comparable to the photon-pair one. Ratios $ R_{s,FF}^{\rm TE,TM} $
and $ R_{i,FF}^{\rm TE,TM} $ given in Eqs.~(\ref{B5}) in Appendix
B and quantifying contributions of the signal and idler noise
photon-number densities relatively to the photon-number densities
$ n_s $ and $ n_i $ given in Eq.~(\ref{12}), respectively, are
plotted in Fig.~\ref{fig11}. They are appropriate for the
structure with 11 layers and the joint signal and idler field
composed of the forward-propagating $ {\rm TE} $-polarized signal
and $ {\rm TM} $-polarized idler photons. Despite the low amount
of Ag embedded in the structure ($ 5\times18 $~nm), the numbers of
signal and idler noise photons are comparable to the number of
emitted photon pairs. Comparable values of ratios $ R_{s,FF}^{\rm
TE,TM} $ (1.20 for $ \vartheta_s = 47 $~deg and $\lambda_s = 738
$~nm) and $ R_{i,FF}^{\rm TE,TM} $ (0.97 for $ \vartheta_i = 61
$~deg and $ \lambda_i = 834 $~nm) for the signal noise and idler
noise fields at the corresponding radial emission angles $
\vartheta $ and for the corresponding frequencies $ \omega $
indicate that the numbers of emitted noise photons depend mainly
on the number of photon pairs generated inside the structure. It
is worth to note that the values of ratios $ R_{s,FF}^{\rm TE,TM}
$ and $ R_{i,FF}^{\rm TE,TM} $ increase in the vicinity of
forbidden bands, i.e. in the area with strong back-scattering and
interference (see Fig.~\ref{fig11}).

As discussed in Appendix~B, photons from photon pairs in which
only one photon enters the detection system represent an
additional source of the noise. In the analyzed structure, photon
pairs with a forward-propagating $ {\rm TE} $-polarized signal
photon and a backward-propagating $ {\rm TM} $-polarized idler
photon contribute to the noise in the signal field. On the other
hand, photon pairs with a backward-propagating $ {\rm TE}
$-polarized signal photon and a forward-propagating $ {\rm TM}
$-polarized idler photon are responsible for an additional noise
in the idler field. As the numbers of emitted photon pairs with
different propagation directions are comparable, the numbers of
noise photons constituting these contributions are also
comparable. However, these noise contributions can be eliminated
if multiple coincidence-count detections are measured.

Considerable amount of the noise present in the generated
photon-pair states restricts applicability of such states to the
schemes based on coincidence-count measurements. In these schemes,
a single-photon noise contributes to the measurement only via
random coincidences that are, however, seldom due to the weakness
of the field. Possible applications suitable for photon-pair
states emitted from metal-dielectric layered structures include
quantum cryptography using photon pairs \cite{Gisin2002} or
quantum optical coherence tomography \cite{Abouraddy2002}, to name
few. On the other hand, these states are not suitable for
constructing heralded single-photon sources \cite{Alibard2005}.

\section{Conclusions}
Using quantization of photon flux, a model of spontaneous
parametric down-conversion in metal-dielectric layered structures
has been developed. Applying this model, an efficient structure
composed of six dielectric GaN layers and five metal Ag layers has
been designed and analyzed. Highly enhanced electric-field
amplitudes caused by metal layers not only compensate for losses
in the metal layers, they also allow efficient photon-pair
generation in the nonlinear GaN layers. Despite the small number
of used layers, the generated photon pairs have very narrow
spectra. They are also emitted into very narrow intensity rings in
the transverse plane. Compared to a structure consisting of only
one GaN monolayer with the same amount of material, the analyzed
structure provides photon-pair fluxes greater by seven orders in
magnitude. Correlated areas of the emitted photon pairs are very
narrow and differ for TE- and TM-polarized fields. Whereas they
attain a circular shape for TE-polarized fields and a Gaussian
radially-symmetric transverse pump-beam profile, they are highly
elliptic for TM-polarized fields due to squeezing in the radial
direction. Temporal intensity correlations in a photon pair occur
at the time scale of ns owing to many back-reflections on the
boundaries. Compared to nonlinear dielectric layered structures,
photon-pair fluxes greater by four orders in magnitude are found.
On the other hand, they also generate a single-photon noise
originating in broken photon pairs and having photon fluxes
comparable to those of photon pairs. Metal-dielectric layered
structures provide in general strongly directionally emitted and
spectrally narrow-band photon pairs necessary, e.g., for
quantum-information processing with photons and atoms.

\acknowledgments Support by projects CZ.1.05/2.1.00/03.0058 and
CZ.1.07/2.3.00/20.0017 of M\v{S}MT \v{C}R and P205/12/0382 of GA
\v{C}R are acknowledged. D.J. and J.P. acknowledge support by
project IGA\_PrF\_2014005 of IGA UP Olomouc. J.S. thanks the projects
CZ.1.07/2.3.00/30.0004 and CZ.1.07/2.3.00/20.0058 of M\v{S}MT
\v{C}R.

\appendix
\section{$ \chi^{(2)} $ tensor for metals}

An expression for tensor $\chi^{(2)}_{jkm}$ of nonlinear
susceptibility appropriate for metals containing electrons moving
by the Lorentz force caused by an external pump field
\cite{Larciprete2008} is derived in the Appendix. Position $ {\bf
r}(t) $ of an electron obeys the following equation of motion
\begin{equation}  
\label{A1}
 \frac{d^2{\mathbf{r}(t)}}{dt^2} + \gamma \frac{d{\mathbf{r}(t)}}{dt} =
 -\frac{e}{m} \mathbf{E}(t) - \frac{e}{m}\frac{d{\mathbf{r}(t)}}{dt}
 \times \mathbf{B}(t),
\end{equation}
in which $m$ stands for the electron mass, $\gamma$ is the
collision factor and $e$ denotes the positive elementary charge.
Symbol $\mathbf{E}$ ($\mathbf{B}$) means the electric- (magnetic-)
field amplitude. Vector product is denoted as $ \times $.
Considering mean volume density of electrons $ N $, macroscopic polarization $ {\bf P}(t)
$ is determined by the expression $ -e N {\bf r}(t) $. Equation
(\ref{A1}) can thus be transformed into the following equation for
polarization $ {\bf P}(t) $:
\begin{equation} 
\label{A2}
 \pa{^2\textbf{P}}{t^2} + \gamma \pa{\textbf{P}}{t} =
 \varepsilon_0 \Omega_p^2 \textbf{E} - \frac{e}{m} \pa{\textbf{P}}{t}
 \times \textbf{B};
\end{equation}
$\Omega_p \equiv e \sqrt{N/(\varepsilon_0 m)} $ is the plasma
frequency.

Perturbation approach is applied to find the solution of
Eq.~(\ref{A2}). Polarization $ {\bf P} $ is decomposed into strong
linear and weak nonlinear parts. Also, the second term on the
right-hand side of Eq.~(\ref{A2}) is much smaller than the first
one. Solution of Eq.~(\ref{A2}) for three monochromatic waves
representing the pump, signal and idler fields can then be easily
found following \cite{Boyd2003}. It allows us to express the
nonlinear tensor $ \chi^{(2)} $ as follows:
\begin{eqnarray}  
  \chi^{(2)}_{jlm}(\mathbf{k}_p,\mathbf{k}_s,\mathbf{k}_i) &=&
  -\frac{i \varepsilon_0}{2\pi N e}
  \sum_{o,q=x,y,z} \nonumber \\
  & & \hspace{-15mm} \Bigl[
  L^\ast(\omega_p)L^\ast(\omega_s)A(\omega_s,\omega_i)
  \varepsilon_{jlo}\varepsilon_{oqm} k^\ast_{i,q} \nonumber \\
  & & \hspace{-15mm} \mbox{} + L^\ast(\omega_p)L^\ast(\omega_i)A(\omega_i,\omega_s)
  \varepsilon_{imo} \varepsilon_{oql} k^\ast_{s,q} \nonumber \\
  & & \hspace{-15mm} \mbox{} + L(\omega_i)L^\ast(\omega_s)A(\omega_s,\omega_p)
   \varepsilon_{mlo} \varepsilon_{oqj} k_{p,q} \nonumber \\
  & & \hspace{-15mm} \mbox{} - L(\omega_i)L(\omega_p)A(\omega_p,\omega_s) \varepsilon_{mjo}
   \varepsilon_{oql} k_{s,q}^\ast \nonumber\\
  & & \hspace{-15mm} \mbox{} - L(\omega_s)L(\omega_p)A(\omega_p,\omega_i)
  \varepsilon_{ljo} \varepsilon_{oqm} k_{i,q}^\ast \nonumber\\
  & & \hspace{-15mm} \mbox{} + L(\omega_s)L^\ast (\omega_i)A(\omega_i,\omega_p)
  \varepsilon_{lmo} \varepsilon_{oqj} k_{p,q} \Bigr].
\label{A3}
\end{eqnarray}
In Eq.~(\ref{A3}), $\varepsilon_{ijk}$ denotes the Levi-Civita
tensor, $ L(\omega) = \Omega_p^2 / (\omega^2 + i \gamma \omega) $
and $ A(\omega,\omega') = \omega/\omega' $. The expression in
Eq.~(\ref{A3}) for tensor $ \chi^{(2)} $ reveals its strong
dependence on frequencies of the interacting fields. Wave vectors
$ {\bf k} $ occurring in Eq.~(\ref{A3}) are assumed to be complex,
as the fields are strongly attenuated in metals (due to the skin
effect). The expected values of elements of $\chi^{(2)}$ tensor
for metals are of the order of $10^{-13} $~m/V.

\section{Losses in layered structures and noise photons}

The analyzed metal-dielectric layered structures may produce
considerable amount of noise photons due to strong absorption of
the metal. The reason is that an absorbed photon leaves its twin
in the structure. If this twin exits the structure, it forms the
noise that is superimposed on the emitted photon-pair field. In
this Appendix, we develop a theory that quantifies the
contribution of noise photons. We assume for simplicity that
photon pairs are generated only in dielectric layers, in accord
with our results that have revealed only weak generation of photon
pairs in metal layers. However, the inclusion of metal layers as
sources of photon pairs is straightforward.

Detailed inspection of Eq.~(\ref{8}) for two-photon state $
|\psi^{\rm out}_{s,i}\rangle $ reveals that this state is composed
of contributions describing photon pairs emitted in different
layers. We assume that similar decomposition can be done also for
the joint signal-idler photon-number density $
n^{\alpha\beta}_{ab}({\bf \Omega}_s,{\bf \Omega}_i) $ defined in
Eq.~(\ref{10}):
\begin{eqnarray} 
 n^{\alpha\beta}_{ab}({\bf \Omega}_s,{\bf \Omega}_i) &\approx&
  \sum_{l\in {\rm diel}} \, \sum_{a',b'=F,B}
  T^{(l)\alpha}_{s,aa'}({\bf \Omega}_s) T^{(l)\beta}_{i,bb'}({\bf \Omega}_i)
  \nonumber \\
 & & \mbox{} \times
  n^{(l)\alpha\beta}_{a'b'}({\bf \Omega}_s,{\bf \Omega}_i) .
\label{B1}
\end{eqnarray}
In Eq.~(\ref{B1}), symbol $ n^{(l)\alpha\beta}_{ab}({\bf
\Omega}_s,{\bf \Omega}_i) $ stands for the joint signal-idler
photon-number density of photon pairs emitted in an $ l$-th layer.
Symbol $ \sum_{l\in{\rm diel}} $ means summation over dielectric
layers. The photon-number density $ n^{(l)\alpha\beta}_{ab}({\bf
\Omega}_s,{\bf \Omega}_i) $ is determined along Eq.~(\ref{10})
using a two-photon spectral amplitude $
\phi^{(l)\alpha\beta}_{ab}({\bf \Omega}_s,{\bf \Omega}_i) $
appropriate for the $ l $-th layer. The intensity transmission
coefficients $T_{m,aa'}^{(l)\alpha} $ introduced in Eq.~(\ref{B1})
give the probability that an $ \alpha $-polarized photon in field
$ m $ propagating in direction $ a'$ in an $ l $-th layer leaves
the structure in direction $ a $.

Whereas $ T_{m,Fa'}^{(l)\alpha} + T_{m,Ba'}^{(l)\alpha} = 1$ holds
for dielectric structures, intensity absorption coefficients
$D_{m,a'}^{(l)\alpha} $ are needed in metal-dielectric structures
to generalize this relation:
\begin{eqnarray}  
 & & T_{m,Fa'}^{(l)\alpha} + T_{m,Ba'}^{(l)\alpha} + D_{m,a'}^{(l)\alpha}= 1 ;
  \nonumber \\
 & & \hspace{10mm}  m=s,i; \hspace{3mm} \alpha ={\rm TE, TM};
  \hspace{3mm} a'=F,B .
\label{B2}
\end{eqnarray}
The intensity absorption coefficient $ D_{m,a'}^{(l)\alpha} $
determines the probability that an $ \alpha $-polarized photon
propagating in direction $ a'$ in an $ l $-th layer in field $ m $
is absorbed inside the structure. Using absorption coefficients
$D_{m,a'}^{(l)\alpha} $, the signal noise photon-number density $
d^{\alpha}_{si,a}({\bf \Omega}_s,{\bf \Omega}_i) $ quantifying the
amount of single $ \alpha $-polarized photons at frequency $
\omega_s $ propagating at angle $ (\vartheta_s,\psi_s) $ in
direction $ a $ and originating in pairs with an idler photon with
frequency $ \omega_i $ at angle $ (\vartheta_i,\psi_i) $ is
expressed as follows:
\begin{eqnarray} 
 d^{\alpha}_{si,a}({\bf \Omega}_s,{\bf \Omega}_i) &=&
  \sum_{l\in {\rm diel}} \, \sum_{\beta={\rm TE,TM}} \, \sum_{a',b'=F,B}
   T^{(l)\alpha}_{s,aa'}({\bf \Omega}_s) \nonumber \\
 & & \mbox{} \times D^{(l)\beta}_{i,b'}({\bf \Omega}_i)
   n^{(l)\alpha\beta}_{a'b'}({\bf \Omega}_s,{\bf \Omega}_i) .
\label{B3}
\end{eqnarray}
An overall signal noise photon-number density $
d^{\alpha}_{s,a}({\bf \Omega}_s) $ is then simply determined by
integrating over all possible idler-field frequencies $ \omega_i $
and propagation angles $ (\vartheta_i,\psi_i) $:
\begin{equation} 
 d^{\alpha}_{s,a}({\bf \Omega}_s) = \int_{0}^{\infty} d\omega_i
  \int_{-\pi/2}^{\pi/2} d\vartheta_i \int_{-\pi/2}^{\pi/2}
  d\psi_i \, d^{\alpha}_{si,a}({\bf \Omega}_s,{\bf \Omega}_i).
\label{B4}
\end{equation}
Formulas analogous to those written in Eqs.~(\ref{B3}) and
(\ref{B4}) can be derived also for the idler-field noise
contribution.

To judge contributions of noise single photons to the generated
state with $ \alpha $-polarized signal photons in direction $ a $
and $ \beta $-polarized idler photons in direction $ b $, we
define ratios $ R_{m,ab}^{\alpha\beta}({\bf \Omega}_m) $ of noise
photon-number densities $ d_{s,a}^{\alpha}({\bf \Omega}_s) $ and $
d_{i,b}^{\beta}({\bf \Omega}_i) $ with respect to densities $
n_{m,ab}^{\alpha\beta}({\bf \Omega}_m) $ belonging to photon pairs
and written in Eq.~(\ref{11}):
\begin{equation}   
 R_{s,ab}^{\alpha\beta}({\bf \Omega}_s) = \frac{
  d_{s,a}^{\alpha}({\bf \Omega}_s) }{ n_{s,ab}^{\alpha\beta}({\bf \Omega}_s) },
  \hspace{5mm}
 R_{i,ab}^{\alpha\beta}({\bf \Omega}_i) = \frac{
  d_{i,b}^{\beta}({\bf \Omega}_i) }{ n_{i,ab}^{\alpha\beta}({\bf \Omega}_i)}.
\label{B5}
\end{equation}

Also photon pairs with polarizations and propagation directions
different from the analyzed one and denoted by indices $(a,\alpha)
$ and $ (b,\beta) $ in Eq.~(\ref{B5}) contribute to noise photons
provided that one of their two photons is captured by detectors.
In this case, ratios $ \tilde{R}_{m,ab}^{\alpha\beta}({\bf
\Omega}_m) $ defined along the relations
\begin{eqnarray}   
 \tilde{R}_{s,ab}^{\alpha\beta}({\bf \Omega}_s) &=& \frac{
  d_{s,a}^{\alpha}({\bf \Omega}_s) + \sum_{\beta'={\rm TE}}^{\rm
  TM} \sum_{b'= F}^{B} n_{s,ab'}^{\alpha\beta'}({\bf \Omega}_s)
   }{ n_{s,ab}^{\alpha\beta}({\bf \Omega}_s) } - 1 ,
  \nonumber \\
 \tilde{R}_{i,ab}^{\alpha\beta}({\bf \Omega}_i) &=& \frac{
  d_{i,b}^{\beta}({\bf \Omega}_i) + \sum_{\alpha'={\rm TE}}^{\rm TM}
  \sum_{a'= F}^{B} n_{i,a'b}^{\alpha'\beta}({\bf \Omega}_i) }{ n_{i,ab}^{\alpha\beta}({\bf \Omega}_i)}
  -1 \nonumber \\
 & &
\label{B6}
\end{eqnarray}
appropriately characterize the noise of the emitted state.
However, this part of noise can be removed in principle when
multiple coincidence-count measurements are applied in the
experiment.

\begin{figure}   
 \vspace{0.5cm}
 \includegraphics[scale=0.8]{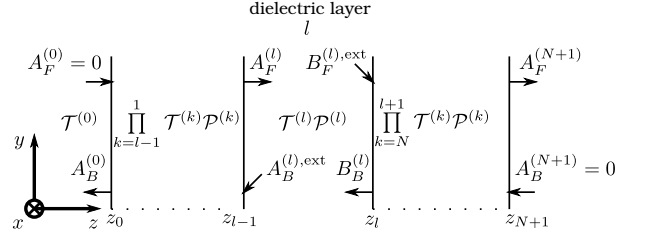}
 \caption{Scheme of a structure composed of $ N $ layers.
  Whereas amplitudes $ A^{(l)} $ describe the fields at the left-hand side
  of an $ l$-th layer ($ l=1,\ldots,N+1 $), amplitudes $ B^{(l)} $ are appropriate for
  the right-hand side of this layer ($ l=1,\ldots,N $);
  amplitudes $ A^{(0)} $ give the fields in front of the structure.
  Amplitudes $ A^{(l),{\rm ext}}_{B} $ and $ B^{(l),{\rm ext}}_{F}
  $ belong to the fields added into an $ l $-th layer.
  Subscript $ F $ ($ B $) identifies the forward- (backward-) propagating fields.
  Matrices $ {\cal T}^{(l)} $ characterize an $ l $-th boundary and matrices $ {\cal P}^{(l)} $
  determine the free-field evolution in an $ l $-th layer.}
 \label{fig12}
\end{figure}

To determine ratios $ R_{m,ab}^{\alpha\beta}({\bf \Omega}_m) $ and
$ \tilde{R}_{m,ab}^{\alpha\beta}({\bf \Omega}_m) $ characterizing
noise in the emitted state, we need intensity transmission and
absorption coefficients for the signal and idler photons born in
each dielectric layer. In what follows, we concentrate our
attention to field $ m $ ($ m=s,i $) and an $ l $-th layer (for
the scheme of a general structure, see Fig.~\ref{fig12}). To
describe properly damping in metal layers, we have to introduce
time into the description, at least implicitly. We reach this by
defining the appropriate boundary conditions. We have to
distinguish two cases characterizing the photons propagating
forward and backward in the $ l $-th layer.

We first add to the $ l $-th layer backward-propagating $
\alpha$-polarized photons described by amplitude $ A^{(l),{\rm
ext}}_{m_B,\alpha}({\bf \Omega}_m) $ and follow their evolution
inside the structure. This evolution is described by the
transfer-matrix formalism elaborated for the nonlinear layered
structures in \cite{PerinaJr2011,PerinaJr2014}. The remaining
boundary conditions are such that photons do not enter the
structure from its front [$ A^{(0)}_{m_F,\alpha}({\bf \Omega}_m)
=0$] and rear [$ A^{(N+1)}_{m_B,\alpha}({\bf \Omega}_m) =0 $]
ends. The backward-propagating photons added into the $ l $-th
layer propagate first in the layers to the left from the $ l $-th
layer, they can penetrate into the layers to the right from the $
l $-th layer later and they can even return back to the
left-hand-side layers from the right-hand-side ones. Following the
scheme plotted in Fig.~\ref{fig12} and showing the used
amplitudes, we can write two sets of linear equations
characterizing the propagation through the left- and
right-hand-side layers separately:
\begin{eqnarray} 
 & &\left(\begin{array}{c} A_{m_F,\alpha}^{(l)}({\bf \Omega}_m) \\
  A_{m_B,\alpha}^{(l),{\rm ext}}({\bf \Omega}_m) + [{\cal P}_{m}^{(l)}({\bf
  \Omega}_m)]_{22}^* B_{m_B,\alpha}^{(l)}({\bf \Omega}_m) \end{array}\right)
  = \nonumber \\
 & & \hspace{35mm} {\cal L}^{(l)}_{m,\alpha}({\bf \Omega}_m)
  \left(\begin{array}{c}
  0 \\ A_{m_B,\alpha}^{(0)} ({\bf \Omega}_m)
  \end{array}\right), \nonumber \\
 & &\left(\begin{array}{c} A_{m_F,\alpha}^{(N+1)} ({\bf \Omega}_m)\\
  0 \end{array}\right)
  = \nonumber \\
 & & \hspace{15mm} {\cal R}^{(l)}_{m,\alpha}({\bf \Omega}_m)
  \left(\begin{array}{c} [{\cal P}_{m}^{(l)}({\bf
  \Omega}_m)]_{11} A_{m_F,\alpha}^{(l)}({\bf \Omega}_m) \\
  B_{m_B,\alpha}^{(l)} ({\bf \Omega}_m) \end{array}\right).
   \nonumber \\
 & &
  \label{B7}
\end{eqnarray}

Matrices $ {\cal L}^{(l)}_{m,\alpha}({\bf \Omega}_m) $ [$ {\cal
R}^{(l)}_{m,\alpha}({\bf \Omega}_m) $] introduced in
Eq.~(\ref{B7}) describe the propagation of both forward- and
backward-propagating fields in the layers positioned to the left
[right] from the $ l $-th layer. They can be expressed in terms of
matrices $ {\cal T}^{(j)}_{m,\alpha}({\bf \Omega}_m) $ and $ {\cal
P}^{(j)}_{m}({\bf \Omega}_m) $ characterizing propagation through
a $ j$-th boundary and free-field propagation in a $ j $-th layer,
respectively:
\begin{eqnarray} 
 {\cal L}^{(l)}_{m,\alpha}({\bf \Omega}_m) &=& \prod_{j=l}^{2}
  \left[ {\cal T}^{(j-1)}_{m,\alpha}({\bf \Omega}_m)
  {\cal P}^{(j-1)}_{m}({\bf \Omega}_m) \right]
  {\cal T}^{(0)}_{m,\alpha}({\bf \Omega}_m) , \nonumber \\
 {\cal R}^{(l)}_{m,\alpha}({\bf \Omega}_m) &=& \prod_{j=N}^{l+1}
  \left[ {\cal T}^{(j)}_{m,\alpha}({\bf \Omega}_m)
  {\cal P}^{(j)}_{m}({\bf \Omega}_m) \right]
  {\cal T}^{(l)}_{m,\alpha}({\bf \Omega}_m) . \nonumber \\
 & &
\label{B8}
\end{eqnarray}
More details including definitions of the elements of matrices $
{\cal T}^{(j)}_{m,\alpha}({\bf \Omega}_m) $ and $ {\cal
P}^{(j)}_{m}({\bf \Omega}_m) $ can be found in
\cite{PerinaJr2011,PerinaJr2014}.

Two sets of equations written in (\ref{B7}) are coupled. These
equations can easily be rearranged such that one linear set of
equations for amplitudes $ A_{m_F,\alpha}^{(N+1)} ({\bf \Omega}_m)
$, $ B_{m_B,\alpha}^{(l)} ({\bf \Omega}_m) $, $
A_{m_F,\alpha}^{(l)}({\bf \Omega}_m) $, and $ A_{m_B,\alpha}^{(0)}
({\bf \Omega}_m) $ characterizing the fields leaving the left- and
right-hand-side layers is obtained:
\begin{eqnarray} 
 & & \left( \begin{array}{c} 0 \\ 1
  \\ 0 \\ 0 \end{array} \right) A_{m_B,\alpha}^{(l),{\rm ext}}({\bf \Omega}_m) = {\cal M}_{m,\alpha}^{(l)}({\bf \Omega}_m)
  \left( \begin{array}{c} A_{m_F,\alpha}^{(N+1)}({\bf \Omega}_m) \\
   B_{m_B,\alpha}^{(l)}({\bf \Omega}_m) \\ A_{m_F,\alpha}^{(l)}({\bf \Omega}_m) \\
   A_{m_B,\alpha}^{(0)}({\bf \Omega}_m) \end{array} \right) ,
   \nonumber \\
 & &
\label{B9}   \\
 & & {\cal M}_{m,\alpha}^{(l)}({\bf \Omega}_m) = \nonumber \\
 & & \hspace{10mm}
  \left( \begin{array}{cccc}
   0 & 0 & -1 & [{\cal L}^{(l)}_{m,\alpha}]_{12} \\
   0 & -[{\cal P}^{(l)}_m]_{22}^* & 0 & [{\cal L}^{(l)}_{m,\alpha}]_{22} \\
   -1 & [{\cal R}^{(l)}_{m,\alpha}]_{12} & [{\cal R}^{(l)}_{m,\alpha}]_{11} [{\cal P}^{(l)}_m]_{11} & 0 \\
   0 & -[{\cal R}^{(l)}_{m,\alpha}]_{22} & -[{\cal R}^{(l)}_{m,\alpha}]_{21} [{\cal P}^{(l)}_m]_{11} & 0
   \end{array} \right). \nonumber \\
 & &
\label{B10}
\end{eqnarray}
The solution of Eqs.~(\ref{B9}) provides the output amplitudes
that determine photon fluxes both inside the $ l$-th layer and
outside the whole layered structure. Their analysis provides us
the needed intensity transmission and absorption coefficients as
follows.

According to the Poynting theorem, time-averaged power $
P^{(l)}_{m_B,\alpha} ({\bf \Omega}_m) $ generated in the $ l $-th
layer by the added field $ A_{m_B,\alpha}^{(l),{\rm ext}} $ is
expressed as follows:
\begin{eqnarray}  
 & & P^{(l)}_{m_B,\alpha}({\bf \Omega}_m) =
  n_m^{(l)}(\omega_m)\cos(\vartheta_m^{(l)}) \nonumber \\
 & & \hspace{10mm} \mbox{} \times \Bigl[ |A_{m_B,\alpha}^{(l),{\rm ext}}({\bf \Omega}_m) + [{\cal P}_{m}^{(l)}({\bf
  \Omega}_m)]_{22}^* B_{m_B,\alpha}^{(l)}({\bf \Omega}_m)|^2 \nonumber \\
 & & \hspace{10mm} \mbox{} + |[{\cal P}_{m}^{(l)}({\bf \Omega}_m)]_{11} A_{m_F,\alpha}^{(l)}({\bf \Omega}_m)|^2
  - |A_{m_F,\alpha}^{(l)}({\bf \Omega}_m)|^2 \nonumber \\
 & & \hspace{10mm} \mbox{}  - |B_{m_B,\alpha}^{(l)} ({\bf
  \Omega}_m)|^2 \Bigr] .
\label{B11}
\end{eqnarray}
This power is partly dissipated both in the left- and
right-hand-side layers and its remaining part leaves the structure
either at its front or rear end. Power $ P^{(l)F}_{m_B,\alpha}
({\bf \Omega}_m) $ [$ P^{(l)B}_{m_B,\alpha} ({\bf \Omega}_m) $]
beyond the rear end [in front] of the structure is determined as
follows
\begin{eqnarray}  
 P^{(l)F}_{m_B,\alpha}({\bf \Omega}_m) &=&
  \cos(\vartheta_m) |A_{m_F,\alpha}^{(N+1)}({\bf \Omega}_m)|^2, \nonumber \\
 P^{(l)B}_{m_B,\alpha}({\bf \Omega}_m) &=&
  \cos(\vartheta_m) |A_{m_B,\alpha}^{(0)}({\bf \Omega}_m)|^2.
\label{B12}
\end{eqnarray}
Power $ P^{(l)D}_{m_B,\alpha} ({\bf \Omega}_m) $ dissipated in the
left- and right-hand-side layers can then be derived from the
conservation law of energy:
\begin{equation}  
 P^{(l)D}_{m_B,\alpha}({\bf \Omega}_m) =  P^{(l)}_{m_B,\alpha}({\bf \Omega}_m)
  - P^{(l)F}_{m_B,\alpha}({\bf \Omega}_m)
  - P^{(l)B}_{m_B,\alpha}({\bf \Omega}_m).
\label{B13}
\end{equation}

If power $ P^{(l)}_{m_B,\alpha}({\bf \Omega}_m) $ equals to that
of one photon per second, the powers $ P^{(l)F}_{m_B,\alpha}({\bf
\Omega}_m) $, $ P^{(l)B}_{m_B,\alpha}({\bf \Omega}_m) $ and $
P^{(l)D}_{m_B,\alpha}({\bf \Omega}_m) $ give in turn intensity
transmission coefficients $ T_{m,FB}^{(l)\alpha}({\bf \Omega}_m) $
and $ T_{m,BB}^{(l)\alpha}({\bf \Omega}_m) $ and intensity
absorption coefficient $ D_{m,B}^{(l)\alpha}({\bf \Omega}_m) $:
\begin{eqnarray}  
 T_{m,aB}^{(l)\alpha}({\bf \Omega}_m) &=& \frac{P^{(l)a}_{m_B,\alpha}({\bf
 \Omega}_m) }{ P^{(l)}_{m_B,\alpha}({\bf \Omega}_m) } , \hspace{3mm} a=F,B ,\nonumber \\
 D_{m,B}^{(l)\alpha}({\bf \Omega}_m) &=& \frac{P^{(l)D}_{m_B,\alpha}({\bf
 \Omega}_m) }{ P^{(l)}_{m_B,\alpha}({\bf \Omega}_m) } .
\label{B14}
\end{eqnarray}

Now, we add to the $ l $-th layer forward-propagating $
\alpha$-polarized photons described by amplitude $ B^{(l),{\rm
ext}}_{m_F,\alpha}({\bf \Omega}_m) $. These photons propagate
first in the right-hand-side layers, they enter into the
left-hand-side layers later and they can propagate back to the
right-hand-side layers again. Also in this case, no photon enters
the structure from its front [$ A^{(0)}_{m_F,\alpha}({\bf
\Omega}_m) =0 $] and rear [$ A^{(N+1)}_{m_B,\alpha}({\bf
\Omega}_m) =0 $] ends. Similarly as for the added
backward-propagating photons, we can write two sets of linear
equations characterizing the propagation through the left- and
right-hand-side layers separately:
\begin{eqnarray} 
 & & \left(\begin{array}{c}
  A_{m_F,\alpha}^{(l)}({\bf \Omega}_m ) \\
   { [{\cal P}_{m}^{(l)}({\bf \Omega}_m )]_{22}^{*} }
   B_{m_B,\alpha}^{(l)}({\bf \Omega}_m)
  \end{array}\right)
  = \nonumber \\
 & & \hspace{40mm} {\cal L}^{(l)}_{m,\alpha}({\bf \Omega}_m)
  \left(\begin{array}{c}
  0 \\ A_{m_B,\alpha}^{(0)} ({\bf \Omega}_m)
  \end{array}\right), \nonumber \\
 & & \left(\begin{array}{c} A_{m_F,\alpha}^{(N+1)} ({\bf \Omega}_m)\\
  0 \end{array}\right)
  = {\cal R}^{(l)}_{m,\alpha}({\bf \Omega}_m) \nonumber \\
 & & \hspace{5mm} \mbox{} \times
  \left(\begin{array}{c} B_{m_F,\alpha}^{(l),{\rm ext}}({\bf \Omega}_m) +
   [{\cal P}_{m}^{(l)}({\bf \Omega}_m)]_{11} A_{m_F,\alpha}^{(l)}({\bf \Omega}_m) \\
   B_{m_B,\alpha}^{(l)} ({\bf \Omega}_m) \end{array}\right).
   \nonumber \\
 & &
  \label{B15}
\end{eqnarray}
Matrices $ {\cal L}^{(l)}_{m,\alpha}({\bf \Omega}_m) $ and $ {\cal
R}^{(l)}_{m,\alpha}({\bf \Omega}_m) $ are defined in
Eqs.~(\ref{B8}). Equations (\ref{B15}) can be transformed into a
linear set of equations for amplitudes $ A_{m_F,\alpha}^{(N+1)}
({\bf \Omega}_m) $, $ B_{m_B,\alpha}^{(l)} ({\bf \Omega}_m) $, $
A_{m_F,\alpha}^{(l)}({\bf \Omega}_m) $, and $ A_{m_B,\alpha}^{(0)}
({\bf \Omega}_m) $ of fields leaving the left- and right-hand-side
layers:
\begin{eqnarray} 
 & & \left( \begin{array}{c} 0 \\ 0
  \\ -{ [{\cal R}^{(l)}_{m,\alpha}({\bf \Omega}_m)]_{11} }
  \\ {[{\cal R}^{(l)}_{m,\alpha}({\bf \Omega}_m)]_{21} }
  \end{array} \right)
  B_{m,\alpha}^{(l),{\rm ext}}({\bf \Omega}_m)
  = \nonumber \\
 & & \hspace{30mm} {\cal M}_{m,\alpha}^{(l)}({\bf \Omega}_m)
  \left( \begin{array}{c} A_{m_F,\alpha}^{(N+1)}({\bf \Omega}_m) \\
   B_{m_B,\alpha}^{(l)}({\bf \Omega}_m) \\ A_{m_F,\alpha}^{(l)}({\bf \Omega}_m) \\
   A_{m_B,\alpha}^{(0)}({\bf \Omega}_m) \end{array} \right) ;
   \nonumber \\
 & &
\label{B16}
\end{eqnarray}
matrix $ {\cal M}_{m,\alpha}^{(l)}({\bf \Omega}_m) $ is defined in
Eq.~(\ref{B10}). The solution of Eqs.~(\ref{B16}) allows us to
determine photon fluxes that give the powers discussed above. For
the forward-propagating photons added into the $ l $-th layer,
power $ P^{(l)}_{m_F,\alpha} ({\bf \Omega}_m) $ given into this
layer by the external field with amplitude $
B_{m_F,\alpha}^{(l),{\rm ext}} $ is derived in the form:
\begin{eqnarray}  
 & & P^{(l)}_{m_F,\alpha}({\bf \Omega}_m) =
  n_m^{(l)}(\omega_m)\cos(\vartheta_m^{(l)}) \nonumber \\
 & & \hspace{10mm} \mbox{} \times \Bigl[ |B_{m_F,\alpha}^{(l),{\rm ext}}({\bf \Omega}_m)
  +  [{\cal P}_{m}^{(l)}({\bf \Omega}_m)]_{11} A_{m_F,\alpha}^{(l)}({\bf \Omega}_m)|^2
  \nonumber \\
 & & \hspace{10mm} \mbox{} + |[{\cal P}_{m}^{(l)}({\bf
  \Omega}_m)]_{22}^* B_{m_B,\alpha}^{(l)}({\bf \Omega}_m)|^2
  - |B_{m_B,\alpha}^{(l)} ({\bf \Omega}_m)|^2 \nonumber \\
 & & \hspace{10mm} \mbox{} - |A_{m_F,\alpha}^{(l)}({\bf \Omega}_m)|^2 \Bigr] .
\label{B17}
\end{eqnarray}
This power can be divided into three parts. Its first part [$
P^{(l)F}_{m_F,\alpha}({\bf \Omega}_m) $] is delivered beyond the
rear end of the structure, whereas its second part [$
P^{(l)B}_{m_F,\alpha} ({\bf \Omega}_m) $] is transferred into the
space in front of the structure. Finally, the third part [$
P^{(l)D}_{m_F,\alpha} ({\bf \Omega}_m) $] dissipates inside the
metal layers. These powers then serve for the determination of
intensity transmission coefficients $ T_{m,FF}^{(l)\alpha}({\bf
\Omega}_m) $ and $ T_{m,BF}^{(l)\alpha}({\bf \Omega}_m) $ and
intensity absorption coefficient $ D_{m,F}^{(l)\alpha}({\bf
\Omega}_m) $. Whereas formulas analogous to those written in
Eqs.~(\ref{B11}) and (\ref{B12}) give powers $
P^{(l)F}_{m_F,\alpha} ({\bf \Omega}_m) $, $ P^{(l)B}_{m_F,\alpha}
({\bf \Omega}_m) $ and $ P^{(l)D}_{m_F,\alpha} ({\bf \Omega}_m) $,
expressions derived from those in Eqs.~(\ref{B13}) provide
coefficients $ T_{m,FF}^{(l)\alpha}({\bf \Omega}_m) $, $
T_{m,BF}^{(l)\alpha}({\bf \Omega}_m) $ and $
D_{m,F}^{(l)\alpha}({\bf \Omega}_m) $.

\bibliography{javurek}
\end{document}